\documentclass[superscriptaddress,showkeys,twocolumn,nofootinbib,longbibliography]{revtex4-1}

\usepackage{amsmath}
\usepackage{amssymb}
\usepackage{graphicx}
\usepackage{epsf}
\usepackage{slashed}
\usepackage{enumitem}
\usepackage[czech,english]{babel}
\usepackage[cp1250]{inputenc}
\usepackage{hyperref}
\usepackage{xcolor}
\usepackage{mathtools}

\usepackage[only,llbracket,rrbracket,llparenthesis,rrparenthesis]{stmaryrd} % for comparison with 4 similar parenthesis symbols
\usepackage{accsupp} % for ensuring the right Unicode codepoint upon pasting

\newcommand{\I}{\ensuremath{\mathrm{i}}}% imaginary unit
\newcommand{\e}{\ensuremath{\mathrm{e}}}% Euler number
\newcommand{\eL}{\mathcal{L}}% Lagrangian
\renewcommand{\d}{\ensuremath{\mathrm{d}}}% differential
\newcommand{\threevector}[1]{\boldsymbol{#1}}% three-vector -- jeste zvazit...
\newcommand{\qm}[1]{``#1''} % quotation marks
\newcommand{\sgn}{\mathop{\rm sgn}\nolimits}

\newcommand{\innerdot}{\ensuremath{\!\cdot\!}}% dot product

\usepackage{bbm} %\mathbbm{1}, \mathbbmss{1}, \mathbbmtt{1}

\usepackage[nodayofweek]{datetime}
\newdateformat{mydate}{\twodigit{\THEDAY}{ }\shortmonthname[\THEMONTH], \THEYEAR}
\usepackage[normalem]{ulem}
\begin{document}

\title{Shapes of magnetic monopoles in effective $SU(2)$ models}

\newcommand{\affiliacePraha}{Institute of Experimental and Applied Physics, \\ Czech Technical University in Prague, Husova~240/5, 110~00 Prague~1, Czech Republic}

\newcommand{\affiliaceOpava}{Research Centre for Theoretical Physics and Astrophysics, Institute of Physics, \\ Silesian University in Opava, Bezru\v{c}ovo n\'{a}m\v{e}st\'{\i}~1150/13, 746~01 Opava, Czech Republic}

\author{Petr Bene\v{s}}
\email{petr.benes@utef.cvut.cz}
\affiliation{Institute of Experimental and Applied Physics, \\ Czech Technical University in Prague, Husova~240/5, 110~00 Prague~1, Czech Republic}
%\affiliation{\affiliacePraha}

\author{Filip Blaschke}
\email{filip.blaschke@fpf.slu.cz}
\affiliation{Institute of Experimental and Applied Physics, \\ Czech Technical University in Prague, Husova~240/5, 110~00 Prague~1, Czech Republic}
\affiliation{Research Centre for Theoretical Physics and Astrophysics, Institute of Physics, \\ Silesian University in Opava, Bezru\v{c}ovo n\'{a}m\v{e}st\'{\i}~1150/13, 746~01 Opava, Czech Republic}
%\affiliation{\affiliacePraha}
%\affiliation{\affiliaceOpava}

\begin{abstract}
We present a systematic exploration of a general family of effective $SU(2)$ models with an adjoint scalar. First, we discuss a redundancy in this class of models and use it to identify seemingly different, yet physically equivalent models. Next, we construct the Bogomol'nyi--Prasad--Sommerfield (BPS) limit and derive analytic monopole solutions. In contrast to the 't~Hooft--Polyakov monopole, included here as a special case, these solutions tend to exhibit more complex energy density profiles. Typically, we obtain monopoles with a hollow cavity at their core where virtually no energy is concentrated;
%instead, 
accordingly, 
most of the monopole's energy is stored in a spherical shell around its core. Moreover, the shell itself can be structured, with several \qm{sub-shells}. A recipe for the construction of these analytic solutions is presented.
\begin{center}
\emph{For Ji\v{r}\'{\i} Ho\v{s}ek on his 80th birthday.}
\end{center}
%\remark{Compiled: \mydate\today, {\currenttime} Correction of some typos, even in equations - perhaps should be submitted to arxiv as v3?}
\end{abstract}


\maketitle

%\tableofcontents

\section{Introduction}

%praise the monopole
Magnetic monopoles are amongst the most important hypothetical particles in contemporary physics. Although they still await experimental detection, there are very strong theoretical reasons for their existence. Monopoles play a major r\^{o}le in such disparate subjects as cosmology, particle physics, and condensed matter physics. Their detection would constitute a fundamental breakthrough as they would provide a completely new window to high-energy physics.

%landscape
In \cite{Blaschke:2022pwq} we have introduced a generalization of Lee--Weinberg's conceptual scheme \cite{Lee:1994sk} and presented a landscape of $U(1)$ gauge theories with a scalar field $\phi$ and massive, complex vector fields $W_\mu$, namely:
\begin{eqnarray}
\label{eq:landscape}
\eL_{U(1)} &=&
- \frac{f_1^2(\phi/v)}{4g^2} F_{\mu\nu}^2
- \frac{\eta(\phi/v)}{2} d_{\mu\nu} F^{\mu\nu}
- \frac{\chi(\phi/v) g^2}{4} d_{\mu\nu}^2
\nonumber \\ && {}
- \frac{f_2^2(\phi/v)}{2} \big|D_\mu W_\nu-D_\nu W_\mu\big|^2
+ m^2(\phi/v) \big|W_\mu\big|^2
\nonumber \\ && {}
+ \frac{1}{2} (\partial_\mu \phi)^2
- \frac{\lambda}{2} (\phi^2-v^2)^2
\,,
\end{eqnarray}
where $d_{\mu\nu} = \I (W_\mu^\dag W_\nu - W_\mu W_\nu^\dag)$ is a dipole-moment tensor. These models represent the most general effective field-theoretical descriptions of a Dirac monopole possessing a finite mass. The r\^{o}le of form-functions $f_{1,2}(\phi/v)$, $\eta(\phi/v)$, $\chi(\phi/v)$ and $m(\phi/v)$ is to provide control over the properties of the \qm{substrate}
%in which is the monopole is placed
in which the monopole is placed
($\phi/v$ being the collective coordinate of the coherent state). In particular:
\begin{itemize}
\item $f_1$ is as field-dependent magnetic susceptibility,
\item $f_2$ describes field-dependent non-linear elastic properties of $W_\mu$ fields,
\item $\eta$ and $\chi$ control the dipole-moment interactions of $W_\mu$,
\item $m^2$ is a field-dependent mass.
\end{itemize}
 
Alternatively, we can think about \eqref{eq:landscape} as an effective action of a more fundamental theory with partially resumed loop diagrams that are embodied in these form-functions. In short, we believe that Eq.~\eqref{eq:landscape} is something that one could typically encounter down the stream of the renormalization group river starting from an unknown, more fundamental theory of magnetic monopoles.\footnote{This viewpoint is strengthened by the appearance of four-derivative terms in the canonical formulation of $SU(2)$ landscape in this paper.}

%Why
The utility of considering such a vast landscape is the following: It is perhaps intuitively clear that properties of classical solutions would not be qualitatively different for all choices of form-functions. Instead, they would fall into a small number of universality classes in which a modification of various field-dependent couplings would result in only quantitative difference, but not in a qualitative change in their properties and behavior. By identifying and studying such universality classes, we can understand something about monopoles even though we do not know the ultimate, correct theory, provided, of course, that the \qm{real} monopole in Nature (if ever we found one) falls into one of these universality classes. 

%This paper
In this paper, we make a first step toward this goal and provide a systematic classification of spherically-symmetric solutions of a particular sub-landscape of \eqref{eq:landscape} that is endowed with $SU(2)$ gauge symmetry. The advantages of this restriction are both practical and utilitarian. Practical, as we can focus our efforts more narrowly and utilitarian, as we can compare our solutions with the canonical 't~Hooft--Polyakov monopole \cite{tHooft:1974kcl, Polyakov:1974ek}, which itself is a particular point of the general $SU(2)$ sub-landscape. Furthermore, we can avoid the troubles with residual singularities that are permanent in $U(1)$ models. However, as we will see, the discussion of whether the solutions must be always singularity-free will be one of the corollaries of this paper.

Magnetic monopoles in extended, exotic, and variously generalized models have been studied before \cite{Ramadhan:2015qku, Bazeia:2018eta, Casana:2012un, Bazeia:2018fhg, Ferreira:2021uhk}. One of the characteristics of spherically symmetric single monopole solutions that seems to be linked to form-functions is the shape of their energy density profiles. We will pay a special attention to them. As already reported by Bazeia \emph{et al.}~\cite{Bazeia:2018fhg}, monopoles can have unexpected shapes (which they dubbed \qm{small} and \qm{hollow}), in comparison with the 't~Hooft--Polyakov's solution. In particular, the appearance of a hollow cavity at the core of the monopole, where virtually no energy is stored, seems to be more typical than not. In this paper, we provide explicit analytic solutions for these hollow monopoles and we discuss the conditions under which the cavity appears and what controls its extent.

The paper is organized as follows. First, in Sec.~\ref{theorynonbps}, we introduce a general class of $SU(2)$ models and discuss its form-invariance. Next, in Sec.~\ref{theorybps}, we construct the Bogomol'nyi--Prasad--Sommerfield (BPS) limit and formulate the corresponding BPS equation of motion and energy density. In Sec.~\ref{sphericalsymm} we specialize to spherical symmetry, i.e., to single-monopole solutions, and present a \qm{recipe} for finding analytic solutions. In the subsequent Sec.~\ref{examples}, we show explicit examples. Finally, in Sec.~\ref{conclusions}, we summarize our results and give an outlook to future directions of research.

\section{General (non-BPS) model}
\label{theorynonbps}

\subsection{The Lagrangian}

We consider the most general $SU(2)$ gauge theory that is quadratic in both the derivatives of gauge fields, $\threevector{F}^{\mu\nu}$, and the derivatives of adjoint (real triplet) scalars, $D^\mu\threevector{\phi}$. Accordingly, the Lagrangian must be a linear combination of four algebraically independent terms $(D^\mu \threevector{\phi})^2$, $(\threevector{\phi} \innerdot D^\mu \threevector{\phi})^2$, $(\threevector{F}^{\mu\nu})^2$ and $(\threevector{\phi} \innerdot \threevector{F}^{\mu\nu} )^2$, with the coefficients being dimensionless and gauge-invariant functions of $\threevector{\phi}$. We find it most convenient to write these terms as
\begin{eqnarray}
\label{lagrangian}
\eL &=& 
\frac{v^2}{2} \bigg[
  f_1^2 \, \bigg(\frac{(D^\mu \threevector{\phi})^2}{\threevector{\phi}^2}-\frac{(\threevector{\phi} \innerdot D^\mu \threevector{\phi})^2}{\threevector{\phi}^4}\bigg)
+ f_3^2 \, \frac{(\threevector{\phi} \innerdot D^\mu \threevector{\phi})^2}{\threevector{\phi}^4}
\bigg]
\nonumber \\ && {}
- \frac{1}{4g^2} \bigg[
  f_2^2 \, \bigg((\threevector{F}^{\mu\nu})^2-\frac{(\threevector{\phi} \innerdot \threevector{F}^{\mu\nu} )^2}{\threevector{\phi}^2}\bigg)
+ f_4^2 \frac{(\threevector{\phi} \innerdot \threevector{F}^{\mu\nu})^2}{\threevector{\phi}^2}
\bigg]
\nonumber \\ && {}
- V(\threevector{\phi}^2)
\,,
\end{eqnarray}
where $f_{1,2,3,4}^2$ are the \qm{form-functions} and where
\begin{eqnarray}
D^\mu \threevector{\phi} &=& \partial^\mu \threevector{\phi} + \threevector{A}^\mu \times \threevector{\phi} \,,
\\
\threevector{F}^{\mu\nu} &=& \partial^\mu \threevector{A}^\nu - \partial^\nu \threevector{A}^\mu + \threevector{A}^\mu \times \threevector{A}^\nu \,,
%\\
%V(\threevector{\phi}^2) &=& \frac{1}{4}\lambda \big(\threevector{\phi}^2-v^2\big)^2 \,,
\end{eqnarray}
with the boldface denoting a three-vector in the gauge space. The potential $V(\threevector{\phi}^2) \geq 0$ need not be specified, besides the assumption that for $\threevector{\phi}^{2} = v^2$ (where $v^2 > 0$) it vanishes, triggering the spontaneous symmetry breakdown $SU(2)\to U(1)$.

In the bulk of this paper, we utilize the decomposition
\begin{eqnarray}
\label{definitionvHn}
\threevector{\phi} &=& v H \threevector{n} \,,
\end{eqnarray}
where the isovector $\threevector{n}$ is normalized as $\threevector{n}^2 = 1$ and $H$ is a dimensionless gauge-invariant scalar function, that serves as an argument of all form-functions:
\begin{eqnarray}
f_i^2 &\equiv& f_i^2(H)\,.
\hspace{10mm}
(i=1,2,3,4)
\end{eqnarray}

The link of the $SU(2)$ Lagrangian \eqref{lagrangian} to the $U(1)$ Lagrangian \eqref{eq:landscape} is made by rewriting the former in the unitary gauge and matching the respective form-functions (keeping in mind that $f^2_{1,2}$ in both Lagrangians can be different). At this point, there is no need to make this link explicit and we do not display it here.

Note that the general model \eqref{lagrangian} includes the usual renormalizable Georgi--Glashow model as the special case $f_1^2 = f_3^2 = H^2$ and $f_2^2 = f_4^2 = 1$.

There are two reasons for writing the Lagrangian \eqref{lagrangian} with the basis of projector-like structures. The first one is that the condition for the energy density to be bounded from below is very simple: Each of the corresponding functions $f_i^2$ must be non-negative,
\begin{equation}
f_{i}^2 \geq 0 \,,
\end{equation}
as indicated by the square notation.

The second reason is that the form of the Lagrangian \eqref{lagrangian} is well suited for the formalism to be used later in this paper. E.g., using the decomposition \eqref{definitionvHn} the first two terms in \eqref{lagrangian} (proportional to $f_1^2$ and $f_3^2$) can be respectively rewritten very simply as
\begin{subequations}
\label{TermsPropToh14}
\begin{eqnarray}
\frac{(D^\mu \threevector{\phi})^2}{\threevector{\phi}^2} 
- \frac{(\threevector{\phi} \innerdot D^\mu \threevector{\phi})^2}{\threevector{\phi}^4} &=& (D^\mu \threevector{n})^2
\,,
\\
\frac{(\threevector{\phi} \innerdot D^\mu \threevector{\phi})^2}{\threevector{\phi}^4} &=& \frac{(\partial^\mu H)^2}{H^2}
\,.
\end{eqnarray}
\end{subequations}
After introducing spherically symmetric Ansatz later on, similar simplification will occur also for the gauge part.

Finally, we will for convenience assume, without loss of generality,\footnote{This normalization can be always enforced by redefining $\threevector{\phi} \to \threevector{\phi} / f_1(1)$, $v \to v / |f_1(1)|$
%, $\lambda \to \lambda \, f_1^4(1)$ 
and $g \to g \, |f_2(1)|$, which is equivalent to $f_{1,2} \to f_{1,2}/f_{1,2}(1)$.} the normalization
\begin{eqnarray}
\label{hinormalization}
f_1^2(1) \ = \ f_2^2(1) &=& 1 \,.
\end{eqnarray}
In this way, the kinetic terms are properly normalized in the vacuum $\threevector{\phi}^2 = v^2$.

To be on the safe side, we may require the functions $f_i^2$ to vanish sufficiently fast as $H \to 0$ so that the Lagrangian \eqref{lagrangian} is well defined in the entire range $H \in (-\infty,\infty)$. However, this turns out to be an unnecessary condition in certain special cases. In fact, in Sec.~\ref{examplepower} we present a family of models that contains negative powers of $H$, but that can be related to another, yet physically equivalent family of models via field-redefinition (introduced in the following Sec.~\ref{forminvariance}) which is well defined everywhere.

%However, this requirement seems to be too strong, because, at least in some cases, the Lagrangian can be transformed, using the transformation introduced in the following Sec.~\ref{forminvariance}, into another, yet physically equivalent Lagrangian with only positive powers of $H$. We will present an explicit example in Sec.~\ref{examplepower}.

The non-canonical model \eqref{lagrangian} has necessarily the same spectrum of fluctuations near the $SU(2)$-breaking vacuum as does the Georgi--Glashow model. Namely, there is a massless gauge field (photon) corresponding to the unbroken $U(1)$ subgroup, a pair of massive charged vector fields ($W^\pm$), and a massive real scalar field (Higgs). The latter non-vanishing masses are given, due to the normalization \eqref{hinormalization}, by the standard formulas $M_{W} = vg$ and $M_{\mathrm{Higgs}} = 2v\sqrt{V^{\prime\prime}(v^2)}$.

\subsection{Form-invariance and redundancy}
\label{forminvariance}

\newcommand{\HtoH}{\ensuremath{\alpha}}

Let us consider a transformation (field redefinition) of the scalar triplet $\threevector{\phi}$ to a new $\tilde{\threevector{\phi}}$ that only changes its length, namely
\begin{subequations}
\label{transfomationPhiToTildePhi}
\begin{equation}
\threevector{\phi} = vH\threevector{n}
\hspace{8mm}
\longrightarrow
\hspace{8mm}
\tilde{\threevector{\phi}} = v\tilde{H}\threevector{n}
\,,
\end{equation}
where $\tilde{H}$ is related to $H$ as
\begin{equation}
\label{transfomationHftildeH}
H = \HtoH(\tilde{H}) \,.
\end{equation}
\end{subequations}
Here $\HtoH(\tilde{H})$ is an arbitrary invertible and differentiable function that satisfies $\HtoH^2(\pm 1) = 1$ to preserve the vacuum value. Under this transformation, the original Lagrangian $\eL$ of Eq.~\eqref{lagrangian} transforms into a new Lagrangian $\tilde\eL$ of the form
\begin{eqnarray}
\label{lagrangianTransf}
\tilde\eL &=& 
\frac{v^2}{2} \bigg[
\tilde f_1^2 \bigg(\frac{(D^\mu \tilde{\threevector{\phi}})^2}{\tilde{\threevector{\phi}}^2} - \frac{(\tilde{\threevector{\phi}} \innerdot D^\mu \tilde{\threevector{\phi}})^2}{\tilde{ \threevector{\phi}}^4}\bigg)
+ \tilde f_3^2 \frac{(\tilde{\threevector{\phi}} \innerdot D^\mu \tilde{\threevector{\phi}})^2}{\tilde{\threevector{\phi}}^4}
\bigg]
\nonumber \\ && {}
- \frac{1}{4g^2} \bigg[
  \tilde f_2^2 \bigg((\threevector{F}^{\mu\nu})^2 - \frac{(\tilde{\threevector{\phi}} \innerdot \threevector{F}^{\mu\nu})^2}{\tilde{\threevector{\phi}}^2}\bigg)
+ \tilde f_4^2 \frac{(\tilde{\threevector{\phi}} \innerdot \threevector{F}^{\mu\nu})^2}{\tilde{\threevector{\phi}}^2}
\bigg]
\nonumber \\ && {}
- \tilde V(\tilde{\threevector{\phi}}^2)
\,,
\end{eqnarray}
where $\tilde V(\tilde{\threevector{\phi}}^2) = V(\threevector{\phi}^2)$ and where the new functions $\tilde f_i^2$ are given in terms of the old functions $f_i^2$ as
\begin{eqnarray}
\label{transformationhi}
\tilde f_i^2(\tilde H) &=&
\begin{cases}
f_i^2\big(\HtoH(\tilde H)\big) \,, & (i=1,2,4) \\[2pt]
\displaystyle
f_i^2(\HtoH(\tilde H)\big) \bigg(\tilde{H}\frac{\HtoH^\prime(\tilde H)}{\HtoH(\tilde H)}\bigg)^2 \,, & (i=3)
\end{cases}
\hspace{10mm}
\end{eqnarray}
The non-trivial transformation law for $f_3^2$ is due to the fact that only this term contains a derivative of $H$, as can be seen from \eqref{TermsPropToh14}.

There is a special worth noting case of the transformation \eqref{transfomationPhiToTildePhi}: $H = 1 / \tilde{H}$. Obviously, when this transformation is applied twice, we obtain the original theory, hence we can call it a \emph{duality}. Moreover, since $\tilde H \HtoH^{\prime} / \HtoH = -1$, all four $f_i^2$ transform uniformly as
$\tilde f_i^2(\tilde H) = f_i^2(1/\tilde H)$.
%i.e., $f_3^2$ is not singled-out as in \eqref{transformationhi}.
%Invariance of the theory under the duality \eqref{duality} implies, e.g., that a possible singularity of $H$ can be cured by transformation $H \to 1/H$ (assuming $H$ have no zeros). We will showcase particular examples of this in Sec.~\ref{examples}.

Notably, the transformed Lagrangian $\tilde\eL$ is \emph{of the same form} as $\eL$. We can thus say that $\eL$ is \emph{form-invariant} under the transformation \eqref{transfomationPhiToTildePhi} in the sense that \eqref{transfomationPhiToTildePhi} does not introduce a new \emph{kind} of term that was absent beforehand.
%\footnote{By this, we simply mean that the transformation \eqref{transfomationPhiToTildePhi} does not introduce a new \emph{kind} of term that was absent beforehand.}
Moreover, both $\eL$ and $\tilde\eL$ describe the same physics. This means, in particular, that while the corresponding solutions  can differ, they must both yield the same energy density. Consequently, of the four seemingly independent functions $f_i^2$ only three are (physically) independent, since the other one can be fixed by \eqref{transformationhi}. 
%In this sense the Lagrangian \eqref{lagrangian} can be said to be redundant.

The form-invariance of the theory implies that a singularity in $H$ is not necessarily an issue, as it may be cured by a suitable transformation \eqref{transfomationPhiToTildePhi}. We will showcase a particular example of this in Sec.~\ref{examplepower}.

\subsection{Dressed vs.~canonical fields}
\label{dressedcannonical}

As mentioned, we can utilize the transformation \eqref{transfomationPhiToTildePhi} to make one of the four functions $\tilde f_i^2$ to attain a desired form. Although we shall not adopt any single one, let us point out a few choices of interest.

If, for instance, we want the kinetic term of the scalar triplet $\threevector{\phi}$ to be canonically normalized, we should demand $f_1^2 = H^2$. Another, slightly less obvious choice, is to demand $f_3^2 = H^2$, which makes the kinetic term for the scalar singlet $vH$ canonical.

\newcommand{\threevectorRedef}[1]{\tilde{\threevector{#1}}\vphantom{#1}}
%\remark{Typo: corrected $\tilde{\threevector{A}}^\mu$ to  $\threevectorRedef{A}^\mu$}

To achieve canonical normalization of gauge fields, however, the transformation \eqref{transfomationPhiToTildePhi} is not enough, as it cannot arrange $f_2^2 = 1$. In fact, to make $f_2^2 = 1$ we need to perform a \emph{field-redefinition of the gauge fields} as\footnote{We present a generalization of this transformation, also in combination with the transformation \eqref{transfomationPhiToTildePhi}, in Appendix~\ref{AppendixFieldRedef}.}
\begin{eqnarray}
\label{transAtotildeAspec}
\threevector{A}^\mu &=&
\threevectorRedef{A}^\mu
+ \bigg(\frac{1}{f_2}-1\bigg) \frac{\threevector{\phi} \times \tilde{D}^\mu \threevector{\phi}}{\threevector{\phi}^2}
\,,
\end{eqnarray}
where $\tilde{D}^\mu \threevector{\phi} = \partial^\mu \threevector{\phi} + \threevectorRedef{A}^\mu \times \threevector{\phi}$ is a covariant derivative with respect to the transformed gauge field $\threevectorRedef{A}^\mu$. Accordingly, the Lagrangian \eqref{lagrangian} turns into
\begin{eqnarray}
\label{tildeLafteronlyA}
\tilde\eL &=& 
\frac{v^2}{2} \bigg[
\frac{f_1^2}{f_2^2} \bigg( \frac{(\tilde D^\mu \threevector{\phi})^2}{\threevector{\phi}^2} - \frac{(\threevector{\phi} \innerdot \tilde D^\mu \threevector{\phi})^2}{\threevector{\phi}^4}\bigg)
+ f_3^2 \frac{(\threevector{\phi} \innerdot \tilde D^\mu \threevector{\phi})^2}{\threevector{\phi}^4}
\bigg]
\nonumber \\ &&
- \frac{1}{4g^2} \bigg[
(\threevector{F}^{\mu\nu})^2
+ (f_4^2-1) \frac{(\threevector{\phi} \innerdot \threevector{F}^{\mu\nu})^2}{\threevector{\phi}^2}
\bigg]
- V(\threevector{\phi}^2)
\nonumber \\ &&
- \frac{1}{2g^2} \threevector{d}^{\mu\nu} \innerdot \Bigg\{
f_2 \bigg[ 
\tilde{\threevector{F}}_{\mu\nu}
- \frac{(\threevector{\phi} \innerdot \tilde{\threevector{F}}_{\mu\nu})}{\threevector{\phi}^2} \threevector{\phi}
\bigg]
+ f_4^2 \frac{(\threevector{\phi} \innerdot \tilde{\threevector{F}}_{\mu\nu})}{\threevector{\phi}^2} \threevector{\phi}
\Bigg\}
\nonumber \\ &&
- \frac{1}{4g^2} \bigg[
f_2^2 \bigg((\threevector{d}_{\mu\nu})^2 - \frac{(\threevector{\phi} \innerdot \threevector{d}_{\mu\nu})^2}{\threevector{\phi}^2}\bigg)
+ f_4^2 \frac{(\threevector{\phi} \innerdot \threevector{d}_{\mu\nu})^2}{\threevector{\phi}^2}
\bigg]
\,,
\nonumber \\ &&
\end{eqnarray}
where we denoted (here, the prime represents derivative with respect to $H$)
\begin{eqnarray}
\threevector{d}_{\mu\nu} &\equiv&
\phantom{+\,\,}
H \bigg(\frac{1}{f_2}\bigg)^{\prime} \frac{(\threevector{\phi} \innerdot \tilde{D}_\mu \threevector{\phi})(\threevector{\phi} \times \tilde{D}_\nu \threevector{\phi}) - (\mu\leftrightarrow\nu)}{\threevector{\phi}^4}
\nonumber \\ && {}
- 2 \bigg(\frac{1}{f_2}-1\bigg)
\frac{(\threevector{\phi} \innerdot \tilde{D}_\mu \threevector{\phi})(\threevector{\phi} \times \tilde{D}_\nu \threevector{\phi}) - (\mu\leftrightarrow\nu)}{\threevector{\phi}^4}
\nonumber \\ && {}
+ 2 \bigg(\frac{1}{f_2}-1\bigg)
\frac{\tilde{D}_\mu \threevector{\phi} \times \tilde{D}_\nu \threevector{\phi}}{\threevector{\phi}^2}
\nonumber \\ && {}
+ \bigg(\frac{1}{f_2}-1\bigg)^2 \frac{\threevector{\phi} \innerdot (\tilde{D}_\mu \threevector{\phi} \times \tilde{D}_\nu \threevector{\phi})}{\threevector{\phi}^4} \threevector{\phi}
\,.
\end{eqnarray}
Indeed, the transformation \eqref{transAtotildeAspec} did the trick: The gauge kinetic term in \eqref{tildeLafteronlyA} is canonically normalized. Notice, however, that \eqref{transAtotildeAspec} is not form-invariant, and we pay a price of introducing additional interaction terms (those with $\threevector{d}_{\mu\nu}$) that are of order $\sim \mathcal{O}((D\threevector\phi)^4)$.

However, the appearance of these new terms in \eqref{tildeLafteronlyA} has a clear physical interpretation. In the original Lagrangian \eqref{lagrangian}, the form-function $f_2^2$ models a non-linear response of the substrate (described by the scalar triplet $\threevector{\phi}$) to the presence of $SU(2)$ fields. By transforming from the \emph{dressed} fields $\threevector{A}_\mu$ to new \emph{canonical} fields $\tilde{\threevector{A}}_\mu$, this effect does not disappear, but rather manifests itself through the appearance of the dipole-moment tensor interactions.

These interactions, being higher-order in derivatives and non-renormalizable, are normally omitted from the classical action of Yang--Mill--Higgs theory. However, we can view them as effective, partial resummation of quantum corrections at all orders that are modelled here classically via the function $f_2^2$. In other words, the Lagrangian \eqref{tildeLafteronlyA} represents a semi-classical description of theory with charged bosons that have non-trivial electric and magnetic moments.

Since the \emph{canonical} Lagrangian \eqref{tildeLafteronlyA} is far more complicated than the original \emph{dressed} Lagrangian \eqref{lagrangian}, we will use the \qm{dressed} formalism throughout the paper unless stated otherwise. However, the exact correspondence between both Lagrangians shows that the original theory \eqref{lagrangian} is nothing but canonical $SU(2)$ theory with explicit dipole-moment interactions.

\section{BPS model}
\label{theorybps}

\subsection{Derivation}

%Let us consider the energy density of a static\footnote{Strictly speaking, \qm{static} only means $\partial^0 = 0$. The additional assumption $\threevector{A}^0 = 0$ is made in order to obtain magnetic monopoles and not dyons.} configu\-ration ($\partial^0 = \threevector{A}^0 = 0$) of \eqref{lagrangian}:
%Let us henceforth consider static field configurations, $\partial^0 = \threevector{A}^0 = 0$. Since we are interested in magnetic monopoles and not dyons, let us also assume $\threevector{A}^0 = 0$. The energy density corresponding to \eqref{lagrangian} follows as

%Let us henceforth consider static field configurations, $\partial^0 = \threevector{A}^0 = 0$. (The latter condition is in order to obtain magnetic monopoles and not dyons.) The energy density corresponding to \eqref{lagrangian} follows as

Let us henceforth consider static field configurations $\partial^0 = 0$ with $\threevector{A}^0 = 0$. (The latter condition is in order to obtain magnetic monopoles and not dyons.) The energy density corresponding to \eqref{lagrangian} follows as
\begin{eqnarray}
\label{energydensitygeneral}
\mathcal{E} &=&
\frac{v^2}{2} \bigg[f_1^2 \frac{(D^i \threevector{\phi})^2}{\threevector{\phi}^2}
+ (f_3^2-f_1^2) \frac{(\threevector{\phi} \innerdot D^i \threevector{\phi})^2}{\threevector{\phi}^4}\bigg]
\\ \nonumber && {} 
+ \frac{1}{2g^2} \bigg[f_2^2 (\threevector{B}^i)^2
+ (f_4^2-f_2^2) \frac{(\threevector{\phi} \innerdot \threevector{B}^i)^2}{\threevector{\phi}^2}\bigg]
+ V(\threevector{\phi}^2)
\,,
%\nonumber \\ && {}
\end{eqnarray}
where we defined $\threevector{B}^i \equiv -\frac{1}{2} \epsilon^{ijk} \threevector{F}^{jk}$. This can be rewritten identically as
\begin{eqnarray}
\label{Edensxy}
\mathcal{E} &=& 
\frac{1}{2} \bigg[
\frac{f_1}{H} D^i \threevector{\phi}
- \frac{f_2}{g} \threevector{B}^i
+ x\frac{(\threevector{\phi} \innerdot \threevector{B}^i)}{gv^2H^2} \threevector{\phi}
- y\frac{(\threevector{\phi} \innerdot D^i \threevector{\phi})}{v^2H^3} \threevector{\phi}
\bigg]^2
\nonumber \\ && {}
+ \frac{f_1 f_2}{gH} \partial^i (\threevector{\phi} \innerdot \threevector{B}^i)
- \frac{xf_1 + yf_2 - xy}{gH^2} (\partial^i H) (\threevector{\phi} \innerdot \threevector{B}^i)
\nonumber \\ && {}
+ \frac{f_4^2-(x - f_2)^2}{2g^2v^2H^2}
(\threevector{\phi} \innerdot \threevector{B}^i)^2
+ \frac{f_3^2-(y - f_1)^2}{2v^2H^4}
(\threevector{\phi} \innerdot D^i\threevector{\phi})^2
\nonumber \\ && {}
+ V(\threevector{\phi}^2)
\,,
\end{eqnarray}
where $x$, $y$ are arbitrary functions of $H$. We used $D^i \threevector{B}^i = 0$ in order to write $(D^i \threevector{\phi}) \innerdot \threevector{B}^i = \partial^i (\threevector{\phi} \innerdot \threevector{B}^i)$. 

We can now derive the BPS theory in three steps. First, we discard the potential, setting $V = 0$, but keeping the boundary condition $\threevector{\phi}^2 = v^2$. Next, we demand
\begin{equation}
\label{xpoprve}
x \ = \ f_2 + f_4 \,,
\hspace{10mm}
y \ = \ f_1 + f_3 \,,
\end{equation}
so that the terms on the third line in Eq.~\eqref{Edensxy} vanish. Finally, we require
\begin{eqnarray}
\label{xpodruhe}
\bigg(\frac{f_1 f_2}{gH}\bigg)^\prime
&=&
- \frac{xf_1+yf_2-xy}{gH^2}
\,,
\end{eqnarray}
as then the terms on the second line in \eqref{Edensxy} can be combined into a total derivative:
\begin{eqnarray}
\frac{f_1 f_2}{gH} \partial^i (\threevector{\phi} \innerdot \threevector{B}^i)
- \frac{xf_1 + yf_2 - xy}{gH^2} (\partial^i H) (\threevector{\phi} \innerdot \threevector{B}^i)
& &
\nonumber \\ &&
\hspace{-40mm}
\ = \ 
\partial^i \bigg(\frac{f_1 f_2}{gH} \threevector{\phi} \innerdot \threevector{B}^i\bigg) \,.
\end{eqnarray}

Substituting the expressions \eqref{xpoprve} for $x$, $y$ into \eqref{xpodruhe}, we obtain the equation
\begin{eqnarray}
\label{relationBPSfi}
f_3 f_4 &=& H \big(f_1 f_2 \big)^{\prime} \,,
\end{eqnarray}
i.e., a condition to be satisfied by the functions $f_i$ to allow for a BPS theory in the limit of a vanishing potential.

\subsection{Formulation}

The coveted BPS condition \eqref{relationBPSfi} suggests that, for a BPS theory, it may be more convenient to use another parametrization. Namely, let us consider the set of functions $F_i(H)$ defined as
\begin{subequations}
\label{Finf}
\begin{align}
F_1
&= f_1 f_2 \,,
&
F_2
&= f_1/f_2 \,,
\\
H F_3^\prime
&= f_3 f_4 \,,
&
H F_4^\prime
&= f_3/f_4 \,,
\end{align}
\end{subequations}
with the inverse relations
\begin{subequations}
\begin{align}
f_1^2 &=
F_1 F_2 \,,
&
f_2^2 &=
F_1/F_2 \,,
\\
f_3^2 &=
H^2 F_3^\prime F_4^\prime \,,
&
f_4^2 &=
F_3^\prime/F_4^\prime \,.
\end{align}
\end{subequations}
%The prime is the derivative with respect to $H$.
The functions $F_i$ are defined in terms of $f_i$ uniquely up to a sign, therefore the conditions $f_i^2 \geq 0$ for the stability of the theory become
\begin{equation}
\label{sgnFi}
\sgn F_1 = \sgn F_2 \,,
\hspace{10mm}
\sgn F_3^\prime = \sgn F_4^\prime \,.
\end{equation}
Furthermore, we impose the normalization
\begin{equation}
\big|F_i(1)\big| = 1 \,, \hspace{10mm} (i=1,2,3,4)
\end{equation}
which, for $F_{1,2}$, is equivalent to $f_{1,2}^2(1) = 1$, while for $F_{3,4}$ we utilized the freedom to choose the constant of integration in $F_{3,4} = \int F^\prime_{3,4} \, \d H$ any way we like. Finally, let us note that in terms of $F_i$'s the transformation \eqref{transfomationPhiToTildePhi} attains a simpler form (compared with \eqref{transformationhi})
\begin{eqnarray}
\label{transformationFi}
\tilde F_i(\tilde H) &=& F_i\big(\HtoH(\tilde H)\big) \,.
\hspace{8mm}(i=1,2,3,4)
\end{eqnarray}
%i.e., is simpler than in terms of $f_i^2$ -- compare with \eqref{transformationhi}.

The Lagrangian \eqref{lagrangian} (with $V = 0$) can be equivalently expressed as
\begin{eqnarray}
\label{lagrangianF}
\eL &=&
\frac{v^2}{2} \bigg[
F_1 F_2 \frac{(D^\mu \threevector{\phi})^2}{\threevector{\phi}^2}
+ \big(H^2 F_3^\prime F_4^\prime - F_1 F_2 \big) \frac{(\threevector{\phi} \innerdot D^\mu \threevector{\phi})^2}{\threevector{\phi}^4}
\bigg]
\nonumber \\ && {}
- \frac{1}{4g^2} \bigg[
\frac{F_1}{F_2} (\threevector{F}^{\mu\nu})^2
+ \bigg(\frac{F_3^\prime}{F_4^\prime} - \frac{F_1}{F_2}\bigg) \frac{(\threevector{\phi} \innerdot \threevector{F}^{\mu\nu})^2}{\threevector{\phi}^2}
\bigg]
\,.
\end{eqnarray}
Notice that now the renormalizable Georgi--Glashow model corresponds to a special (and rather symmetric) case
$F_1 = F_2 = F_3 = F_4 = H$.

The main benefit of the new parametrization is that the BPS condition \eqref{relationBPSfi} acquires a very simple form
\begin{eqnarray}
\label{relationBPSFi}
F_3 &=& F_1 \,,
\end{eqnarray}
as well as the BPS equation itself:\footnote{The BPS equation is obtained simply as a condition for the square bracket in \eqref{Edensxy} to vanish (so that the energy density is just a total derivative). However, as the resulting equation is a bit messy and complicated, some of its consequences have to be used to obtain the elegant form \eqref{BPSeqnFi}.}
\begin{eqnarray}
\label{BPSeqnFi}
D^i \threevector{\phi}
&=&
\frac{H}{g} \bigg[
\frac{1}{F_2} \bigg(\threevector{B}^i
- \frac{\threevector{\phi} \innerdot \threevector{B}^i}{\threevector{\phi}^2}\threevector{\phi}\bigg)
+ \frac{1}{H F_4^\prime} \frac{\threevector{\phi} \innerdot \threevector{B}^i}{\threevector{\phi}^2} \threevector{\phi}
\bigg]
\,.
\hspace{10mm}
\end{eqnarray}
When it is satisfied, the energy density is just a total derivative, as required:
\begin{eqnarray}
\label{EdensTotDer}
\mathcal{E} &=&
\partial^i \bigg(\frac{F_1}{gH} \threevector{\phi} \innerdot \threevector{B}^i\bigg)
\,,
\end{eqnarray}
but it can be also rewritten in two other equivalent ways as
\begin{subequations}
\begin{eqnarray}
\mathcal{E} &=&
v^2 \bigg[F_1 F_2
\frac{(D^i \threevector{\phi})^2}{\threevector{\phi}^2}
+ \big(H^2 F_1^\prime F_4^\prime-F_1 F_2\big) \frac{(\threevector{\phi} \innerdot D^i \threevector{\phi})^2}{\threevector{\phi}^4} \bigg]
\nonumber \\ &&
\\ &=&
\frac{1}{g^2} \bigg[\frac{F_1}{F_2} (\threevector{B}^{i})^2
+ \bigg(\frac{F_1^\prime}{F_4^\prime}-\frac{F_1}{F_2}\bigg) \frac{(\threevector{\phi} \innerdot \threevector{B}^i)^2}{\threevector{\phi}^2} \bigg]
\,.
\end{eqnarray}
\end{subequations}

\subsection{Magnetic field and mass of a static solution}

Asymptotically, in the vacuum, the field-strength tensor corresponding to the unbroken \qm{electromagnetic} $U(1)$ subgroup is $F^{\mu\nu} \equiv \frac{1}{v} \threevector{\phi}_{\infty} \cdot \threevector{F}^{\mu\nu}$. Therefore, we can define and calculate the asymptotic magnetic field as
\begin{subequations}
\begin{eqnarray}
B^i &\equiv& -\frac{1}{2} \epsilon^{ijk} F^{jk}
\\ &=&
\frac{1}{v} \threevector{\phi}_{\infty} \cdot \threevector{B}^{i}
\\ &=&
\frac{1}{2} H(\infty) \varepsilon^{ijk} (\partial^j\threevector{n}\times\partial^k\threevector{n}) \innerdot \threevector{n}
\,,
\end{eqnarray}
\end{subequations}
where $H(\infty) = \pm 1$ is value of $H$ at spatial infinity. The magnetic charge of a static configuration follows as
\begin{subequations}
\label{qm}
\begin{eqnarray}
q_{\mathrm{m}} &=& \lim_{r \to \infty} \int_{S^2} \d S_i \, B_i
\\ &=& \frac{1}{2} H(\infty)
%\smash{
\underbrace{
\lim_{r \to \infty} \int_{S^2} \d S_i \,  \varepsilon^{ijk} (\partial^j\threevector{n}\times\partial^k\threevector{n}) \innerdot \threevector{n}
}_{\mathclap{\displaystyle
%\smash{
8 \pi N
%}
}}
%}
\hspace{10mm}
\\ &=& 
H(\infty) 4 \pi N
\,,
\end{eqnarray}
\end{subequations}
where $N \in \mathbb{Z}$ is the degree of the mapping $\threevector{n}$.

We calculate the mass of a static configuration using the Gauss--Ostrogradsky theorem (valid for regular $\mathcal{E}$). Using the previous result \eqref{qm}, we obtain
\begin{subequations}
\label{Mgeneral}
\begin{eqnarray}
M &=& \int_{\mathbb{R}^3} \d^3 x \, \mathcal{E}
\\ &=& \int_{\mathbb{R}^3} \d^3 x \, \partial^i \bigg(
\frac{F_1}{gH}
\threevector{\phi} \innerdot \threevector{B}^i\bigg)
\\ &=& \lim_{r \to \infty} \int_{S^2} \d S_i \,
\frac{F_1}{gH}
\threevector{\phi} \innerdot \threevector{B}^i
\\ &=&
\frac{vF_1(1)}{2g}
\underbrace{
\lim_{r \to \infty} \int_{S^2} \d S_i \,
\varepsilon^{ijk} (\partial^j\threevector{n}\times\partial^k\threevector{n}) \innerdot \threevector{n}
}_{\mathclap{\displaystyle
8 \pi N
= 2 H(\infty) q_{\mathrm{m}}
}}
\hspace{10mm}
\\ &=&
F_1(1)
\frac{4 \pi v}{g} N
\,,
\end{eqnarray}
\end{subequations}
where $F_1(1) = \pm 1$. It follows that for the mass to be positive, the two signs must be related as
\begin{eqnarray}
\label{condf1f2N}
F_1(1) \, \sgn N &=& +1 \,,
\end{eqnarray}
so that $M = \frac{4 \pi v}{g} |N| \geq 0$.

Moreover, the magnetic charge can be rewritten as $q_{\mathrm{m}} = \sigma \, 4 \pi |N|$, where the sign $\sigma \equiv F_1(1) H(\infty) = \pm 1$ informs us whether we have a monopole ($q_{\mathrm{m}} > 0$) or antimonopole ($q_{\mathrm{m}} < 0$).

\section{Spherical symmetry}
\label{sphericalsymm}

\subsection{The Ansatz}

Let us now focus on a spherically symmetric $N=1$ solutions for which we adopt the standard \qm{hedgehog} Ansatz:
%\begin{subequations}
%\label{ansatz}
%\begin{eqnarray}
%\threevector{\phi} &=& v H \frac{\threevector{r}}{r} \,,
%\\
%\threevector{A}^i &=& 
%- \frac{\threevector{r} \times \partial^i \threevector{r}}{r^2} (1-K)
%\,,
%\hspace{10mm}
%\end{eqnarray}
%\end{subequations}
%or in components,
%\begin{subequations}
%\label{ansatzcomponents}
%\begin{eqnarray}
%\phi_a &=& v H \frac{x_a}{r} \,,
%\\
%A^i_a &=& -\frac{1}{r^2} \varepsilon_{abi} x_b (1-K) \,.
%\end{eqnarray}
%\end{subequations}
\begin{equation}
\label{ansatz}
\threevector{\phi} \ = \ v H \frac{\threevector{r}}{r} \,,
\hspace{10mm}
\threevector{A}^i \ = \
- \frac{\threevector{r} \times \partial^i \threevector{r}}{r^2} (1-K)
\,,
\end{equation}
or in components,
\begin{equation}
\label{ansatzcomponents}
\phi_a \ = \ v H \frac{x_a}{r} \,,
\hspace{10mm}
A^i_a \ = \ -\frac{1}{r^2} \varepsilon_{abi} x_b (1-K) \,.
\end{equation}

Both form factors $H$ and $K$ are functions of $r = |\threevector{r}|$ and satisfy the  boundary conditions
\begin{equation}
\label{boundarycondHK}
H(\infty) \ = \ 1 \,,
\hspace{10mm}
K(\infty) \ = \ 0 \,,
\end{equation}
that follow from the requirement of the finite total energy. Notice that we have deliberately chosen $H(\infty) = + 1$ instead of $-1$; this, together with
\begin{eqnarray}
\label{f1f2sign}
F_1(1) \ = \ F_2(1) &=& +1 \,,
\end{eqnarray}
that follows from \eqref{condf1f2N} and from $N = 1$, means that we are considering, without loss of generality, only monopoles and not anti-monopoles.

At this point it is customary to impose the conditions 
%\begin{equation}
%H(0) \ = \ 0 \,,
%\hspace{10mm}
%K(0) \ = \ 1 \,, 
%\end{equation}
$H(0) = 0$ and $K(0) = 1$, 
so that the fields $\threevector{\phi}$, $\threevector{A}^i$ are regular at the origin. (Provided that $K-1$ and $H$ go to zero fast enough, i.e., at least linearly and quadratically, respectively.) In case of 't~Hooft--Polakov monopole in the Georgi--Glashow theory (i.e., $F_{i} = H$ for all $i$), these conditions are critical for having a finite mass. However, in our enlarged settings (i.e., $F_{i} \neq H$ for some $i$) we should carefully reconsider their necessity and explore the implications of allowing certain types of singularities in these fields.

Thus, in this paper, we shall adopt a philosophy that a divergence in \emph{gauge-dependent} fields $\threevector{A}^i$ and \emph{form-dependent} field $H$ (see our comment at the end of Sec.~\ref{forminvariance}) is not -- by itself -- necessarily a problem. What will bother us, however, will be regularity of the energy density in the origin, which is without any doubt a physical quantity that must not diverge.

\subsection{The BPS equations and the energy density}

The BPS equation \eqref{BPSeqnFi} under the Ansatz \eqref{ansatz} decouples into a system of two ordinary differential equations for $K$ and $H$:
\begin{subequations}
\label{logKH}
\begin{eqnarray}
\label{logK}
\partial_\rho(\log K) &=& - F_2(H)
\,,
\\
\label{logH}
\partial_\rho H &=& 
\frac{1-K^2}{\rho^2}
\frac{1}{F_4^\prime(H)}
\,,
\end{eqnarray}
\end{subequations}
where we introduced a  dimensionless radius
\begin{eqnarray}
\rho &\equiv& vgr \,.
\end{eqnarray}

%One immediate consequence of the BPS equations is the asymptotic behavior of $K$. From \eqref{logK} we have
From the first equation we can immediately read off
%(due to $H(\infty) = 1$ and $F_2(1) = 1$)
(due to \eqref{boundarycondHK} and \eqref{f1f2sign})
the asymptotic behavior of $K$:
%\begin{eqnarray}
%\label{KinftyExp}
%\partial_\rho(\log K) &\xrightarrow[\rho \to \infty]{}&
%- F_2(1) \ = \ 
%-1 \,,
%\end{eqnarray}
%so that $K$ must for large $\rho$ behave like an exponential:
\begin{eqnarray}
K &\sim& \exp(-\rho)
\hspace{5mm}
\mbox{as}
\hspace{5mm}\rho \to \infty \,.
\end{eqnarray}

The energy density reads 
\begin{eqnarray}
\frac{\mathcal{E}}{v^4g^2} &=&
\frac{\partial_\rho \big[F_1 (1-K^2) \big]}{\rho^2}
%\ = \ 
\\ &=&
\label{EdensitySphericalFi}
2 F_1 F_2 \frac{K^2}{\rho^2}
+
\frac{F_1^\prime}{F_4^\prime}
\frac{(1-K^2)^2}{\rho^4}
\,,
\end{eqnarray}
where, on the second line, we used the BPS equations to eliminate the derivatives with respect to $\rho$. The first line can be immediately integrated over 3-volume to obtain the mass of the monopole 
\begin{eqnarray}
M &=& \int_{\mathbb{R}^3}\!\d^3 x \, \mathcal{E}
\ = \ \frac{4 \pi v}{g} \int_{0}^{\infty}\!\d\rho \, \rho^2 \frac{\mathcal{E}}{v^4 g^2}
\ = \ 
\frac{4 \pi v}{g}
\,,
\hspace{10mm}
\end{eqnarray}
in agreement with the general formula \eqref{Mgeneral} (using $N = 1$ and \eqref{f1f2sign}).

\subsection{Solving the BPS equations}
\label{InvertibleCaseGeneral}

\newcommand{\mnDlogK}{\ensuremath{\kappa}}

We can recast the system of the two first-order equations \eqref{logKH} into a single second-order equation by a straightforward substitution. 
First, let us for convenience introduce the shorthand
\begin{eqnarray}
\label{H0_InvertibleCase}
\mnDlogK &\equiv& - \partial_\rho(\log K) \,.
\end{eqnarray}
Also, from now on we assume $F_2$ to be invertible.\footnote{However, see the discussion at the end of the paper in Sec.~\ref{conclusions}.}
Then we have
\begin{eqnarray}
\label{HfromF2}
H &=&
%F_2^{-1}(H_0) = 
F_2^{-1}(\mnDlogK) \,.
\end{eqnarray}
Next, we introduce two functions $F$, $G$ of $\mnDlogK$ as
\begin{subequations}
\label{defFG}
\begin{eqnarray}
F(\mnDlogK) &\equiv& F_1\big(F_2^{-1}(\mnDlogK)\big) \,, \\
\label{defG}
G(\mnDlogK) &\equiv& F_4\big(F_2^{-1}(\mnDlogK)\big) \,.
\end{eqnarray}
\end{subequations}
%\begin{equation}
%\label{defFG}
%F(\mnDlogK) \ \equiv \ F_1\big(F_2^{-1}(\mnDlogK)\big) \,,
%\hspace{7mm}
%G(\mnDlogK) \ \equiv \ F_4\big(F_2^{-1}(\mnDlogK)\big) \,.
%\hspace{1mm}
%\end{equation}
Their derivatives can be expressed as
\begin{equation}
\label{FGprime}
F^\prime(\mnDlogK) \ = \  \frac{F_1^\prime}{F_2^{\prime}} \,,
\hspace{10mm}
G^\prime(\mnDlogK) \ = \  \frac{F_4^\prime}{F_2^{\prime}} \,,
\end{equation}
where the primes on the right-hand sides are derivatives with respect to $H$, which itself is understood as a function of $\mnDlogK$ via \eqref{HfromF2}.

Expressing $H$ from \eqref{logK} as \eqref{HfromF2} and substituting it into \eqref{logH}, we obtain a single second-order equation for the form factor $K$:
\begin{eqnarray}
\label{eqKinS}
\partial_\rho^2(\log K) &=&
- \frac{1-K^2}{\rho^2} \frac{1}{G^\prime(\mnDlogK)}
\,,
\end{eqnarray}
where we used also \eqref{FGprime}.

%The reason we bothered ourselves with defining yet another set of new functions $F$ and $G$ is not only that the latter enters the equation \eqref{eqKinS} for $K$. Consider 
Using the definitions above we can rewrite the energy density \eqref{EdensitySphericalFi} as
\begin{eqnarray}
\label{EdensityInv}
\frac{\mathcal{E}}{v^4g^2} &=&
2 F(\mnDlogK) \mnDlogK \frac{K^2}{\rho^2}
+ \frac{F^\prime(\mnDlogK)}{G^\prime(\mnDlogK)} \frac{(1-K^2)^2}{\rho^4}
\,.
\end{eqnarray}
In this form, the energy density depends only on $K$ (and its derivative via $\mnDlogK$) and there is no explicit dependence on $H$. Therefore the invariance of the energy density under the transformation $H \to \tilde{H}$, which we showed in Sec.~\ref{dressedcannonical}, is now manifest. Also note that while the energy density in the form \eqref{EdensitySphericalFi} depended on three functions, now it depends only on two. Recall that of the original four functions $F_i$ that define our theory, one is removed by the BPS condition and another is removed by the form-invariance of the Lagrangian. The two functions $F$ and $G$ are exactly the two remaining \emph{physical} functions invariant under the transformation \eqref{transfomationPhiToTildePhi}: $F \equiv F_1 \circ F_2^{-1} = F_1 \circ \HtoH \circ \HtoH^{-1} \circ F_2^{-1}$ (and analogously for $G$).

This suggests to switch from the parametrization of our BPS Lagrangian in terms of three functions $F_1$, $F_2$, $F_4$ to another parametrization in terms of $F$ and $G$ and one non-physical function that would merely represent the freedom under the form-invariance \eqref{transfomationPhiToTildePhi}. The most straightforward way to do this is to trade $F_1$, $F_4$ for $F$, $G$ by defining
\begin{subequations}
\label{eq:fandg}
\begin{eqnarray}
F_1(H) &=& F\big(F_2(H)\big) \,, \\
F_4(H) &=& G\big(F_2(H)\big) \,.
\end{eqnarray}
\end{subequations}
%\begin{equation}
%\label{eq:fandg}
%F_1(H) \ = \ F\big(F_2(H)\big) \,,
%\hspace{8mm}
%F_4(H) \ = \ G\big(F_2(H)\big) \,.
%\hspace{1mm}
%\end{equation}
%\begin{equation}
%\label{eq:fandg}
%F_1(H) \ = \ F\big(F_2(H)\big) \,,
%\hspace{7mm}
%F_4(H) \ = \ G\big(F_2(H)\big) \,.
%\hspace{3mm}
%\end{equation}
Here, we let $F_2$ play the r\^{o}le of the non-physical function. In this way we can recast the Lagrangian \eqref{lagrangianF} -- yet again --  as
\begin{eqnarray}
\label{lagrangianFG}
\eL &=& 
\frac{v^2}{2} \bigg[
F F_2 \frac{(D^\mu \threevector{\phi})^2}{\threevector{\phi}^2}
+ \big[(H F_2^\prime)^2 F^\prime G^\prime - F F_2\big] \frac{(\threevector{\phi} \innerdot D^\mu \threevector{\phi})^2}{\threevector{\phi}^4}
\bigg]
\nonumber \\ && {}
- \frac{1}{4g^2} \bigg[
\frac{F}{F_2} (\threevector{F}^{\mu\nu})^2
+ \bigg(\frac{F^\prime}{G^\prime} - \frac{F}{F_2}\bigg) \frac{(\threevector{\phi} \innerdot \threevector{F}^{\mu\nu})^2}{\threevector{\phi}^2}
\bigg]
\,,
\end{eqnarray}
where the primes at $F$, $G$ are derivatives with respect to $\mnDlogK = F_2(H)$. The virtue of this (final!)~form is that it separates the functions $F$, $G$ that are physical from the function $F_2$  that is arbitrary and merely represents our freedom to perform form-invariant transformation \eqref{transfomationPhiToTildePhi}.

Lastly, let us comment on a special case of  $F(\mnDlogK) = 1$. (It follows that also $F_1$ must be then a constant function, $F_1(H) = 1$.) There are two consequences. First, the Lagrangian \eqref{lagrangianFG} reduces to
\begin{eqnarray}
\eL &=& 
\frac{v^2}{2} F_2
(D^\mu \threevector{n})^2
- \frac{1}{4g^2} \frac{1}{F_2} \bigg[
(\threevector{F}^{\mu\nu})^2
- \frac{(\threevector{\phi} \innerdot \threevector{F}^{\mu\nu})^2}{\threevector{\phi}^2}
\bigg]
\,,
\hspace{8mm}
\end{eqnarray}
so that the scalar singlet $H$, living now only in $F_2(H)$, becomes a non-dynamical field. Second, the energy density cannot be simultaneously regular and positive. This can be seen by assuming power function behavior of $K$ for small $\rho$ as $K = a \rho^n + \ldots$ The energy density then reads $\frac{\mathcal{E}}{v^4g^2} = - \frac{(K^2)^\prime}{\rho^2} = - 2na^2\rho^{2n-3} + \ldots$ Thus, either $n \geq 3/2$ (or $n \geq 1/2$, if we allow for integrable singularity), so that $\mathcal{E}$ is in the origin regular (but negative), or $n < 0$, implying positive (but singular) $\mathcal{E}$. For these reasons, we will not consider the case $F(\mnDlogK) = 1$.

\subsection{Recipe for finding analytic solutions of BPS monopoles}
\label{cookbook}

At this point, it is the right time to briefly recapitulate the previous developments and formulate a recipe for finding a theory that allows an analytic monopole solution.
\begin{enumerate}
\item The first step is to find a function $G(\mnDlogK)$ such that the equation \eqref{eqKinS} can be solved in a closed form for~$K(\rho)$.

%Alternatively, one can come up with prescribed $K = K(\rho)$ and look for $G$ that solves \eqref{eqKinS}. The first way is more physical in a sense that the model is fixed at the beginning. The other way can be paraphrased as solution `engineering' that often lead to the presence of special functions in the Lagrangian density, although the solution itself can be simple.

Taking into account the definition \eqref{defG}, we deduce that $G(\mnDlogK)$ must be defined and continuous on some neighborhood of $\mnDlogK = 1$. This is to allow for the existence of a function $F_2(H)$ that would not only satisfy the condition $F_2(1) = 1$ (recall \eqref{f1f2sign} and \eqref{HfromF2}), but to be also invertible for all $H \in [0,\infty)$.

\item Next step is to choose a function $F(\mnDlogK)$. There are two requirements that come from the desired physical properties of the energy density. (Notice that in its expression \eqref{EdensityInv} everything except $F$ is already fixed.) First, $F$ must satisfy
\begin{subequations}
\label{conditionsFgeneral}
\begin{eqnarray}
\sgn F(\mnDlogK) &=& \sgn \mnDlogK \,,
\\
\sgn F^\prime(\mnDlogK) &=& \sgn G^\prime(\mnDlogK) \,,
\end{eqnarray}
\end{subequations}
for all $\mnDlogK$ (i.e., for all $\rho \geq 0$), so that both terms in the energy density \eqref{EdensityInv} are non-negative. Second, $F$ must be such that the energy density is regular.

Having $G^\prime$, $F$ and $K$, we have all physical ingredients that allow us, e.g., to draw a plot of the energy density \eqref{EdensityInv}.

\item The last step is choosing the function $F_2(H)$. This has a twofold meaning. Primarily, it allows us to find the remaining form factor $H$ through the equation \eqref{HfromF2}. More generally, however, with $F_2$ at hand, we can write down the full Lagrangian \eqref{lagrangianFG}. Recall that, taking into account the form-invariance \eqref{transfomationPhiToTildePhi}, all invertible functions $F_2$ are equally good and lead to physically equivalent Lagrangians.

\end{enumerate}

\subsection{Solvable case}
\label{F3HF2prime}

To the best of our knowledge, the only case when an analytic solution to the equation \eqref{eqKinS} is known\footnote{This is not entirely correct. In fact, it \emph{is} possible to find many analytic solutions to \eqref{eqKinS} by turning the logic upside down and using a kind of \qm{reverse engineering}: Given a prescribed form $K(\rho)$, we can then look for $G^\prime(\mnDlogK)$ that solves \eqref{eqKinS}. However, as simple as it sounds, it is surprisingly difficult to find in this way a $G^\prime$ that meets the conditions for its domain, as discussed at the point 1.~of Sec.~\ref{cookbook}. In particular, all $G^\prime$ we managed to find using this method happened to be singular at $\mnDlogK = 1$.}
is when %$G$ is the identity function:
\begin{eqnarray}
G(\mnDlogK) &=& \mnDlogK \,.
\end{eqnarray}
This, by the way, corresponds to
\begin{eqnarray}
\label{condF4}
F_4 &=& F_2 \,,
\end{eqnarray}
which is certainly satisfying from an aesthetic point of view as it sort of symmetrizes the theory -- compare it with the BPS condition $F_3 = F_1$.

%Let us now consider the special case
%\begin{eqnarray}
%\label{condF4}
%F_4 &=& F_2 \,.
%\end{eqnarray}
%This condition is certainly satisfying from the aesthetical point of view as it sort of symmetrizes the theory --- compare it with the similar BPS condition $F_3 = F_1$. However, its main virtue is that it is equivalent to
%\begin{eqnarray}
%G(\mnDlogK) &=& \mnDlogK \,.
%\end{eqnarray}
%The point is that the equation \eqref{eqKinS} for $K$ is, upon substituting $G^\prime(\mnDlogK) = 1$ into it, analytically solvable. Moreover, this is, to the best of our knowledge, in fact the only case when the general analytical solution (including constants of integration) to \eqref{eqKinS} is known.

Upon substituting $G^\prime(\mnDlogK) = 1$ into the equation \eqref{eqKinS}, we obtain the general solution $K = c\rho / \sinh[c(\rho-\rho_0)]$, where the constants of integration $c$, $\rho_0$ are in principle arbitrary. After all, this solution satisfies the boundary condition $K(\infty) = 0$ for all values of $c$, $\rho_0$.

To determine the value of $c$, we have to consider the other form factor, $H$, that can be calculated from $F_2(H) = \mnDlogK$, where $\mnDlogK = - \partial_\rho(\log K) = c \coth[c(\rho-\rho_0)] - 1/\rho$. From $\lim_{\rho\to\infty}\mnDlogK = c$ (together with the normalization condition $F_2(1) = 1$) we learn that only $c = 1$ is consistent with the boundary condition $H(\infty) = 1$.

\begin{figure}[t]
\begin{center}
\includegraphics[width=0.5\textwidth]{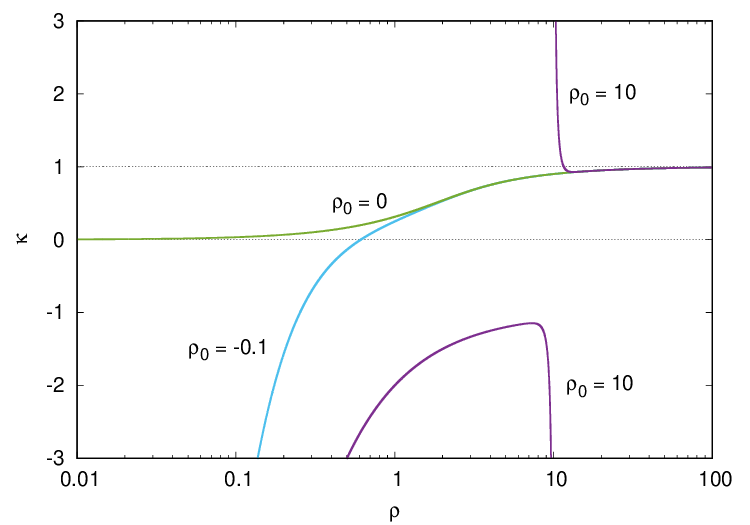}
\caption{Behavior of the function $\mnDlogK = \coth(\rho-\rho_0) - 1/\rho$ for various $\rho_0$. For $\rho_0 < 0$ it smoothly interpolates between $-\infty$ and $1$, while for $\rho_0 = 0$ between $0$ and $1$. For $\rho_0 > 0$ there is a singularity at $\rho = \rho_0$.}
\label{fig_inv_HO}
\end{center}
\end{figure}

%Next, to determine $\rho_0$, we have to resort to the requirement that each of the two terms in the energy density
%\begin{eqnarray}
%\frac{\mathcal{E}}{v^4g^2} &=&
%F(\mnDlogK) \mnDlogK \frac{2K^2}{\rho^2}
%+ F^\prime(\mnDlogK) \frac{(1-K^2)^2}{\rho^4}
%\end{eqnarray}
%has to be separately both regular and non-negative. However, a simple inspection reveals that these conditions cannot be met for any function $F(\mnDlogK)$ as long as $\rho_0 \neq 0$, due to the behavior of $\mnDlogK(\rho)$ in such a case (see Fig.~\ref{fig_inv_HO}). Accordingly, we conclude that only $\rho_0 = 0$ is viable.

Next, to determine $\rho_0$, we have to resort to the requirement that each of the two terms in the energy density
\begin{eqnarray}
\frac{\mathcal{E}}{v^4g^2} &=&
F(\mnDlogK) \mnDlogK \frac{2K^2}{\rho^2}
+ F^\prime(\mnDlogK) \frac{(1-K^2)^2}{\rho^4}
\end{eqnarray}
has to be separately both regular and non-negative. (The latter is nothing but the requirement that each of the terms in \eqref{energydensitygeneral}, proportional to $f_i^2$, be non-negative.)

Let us, then, first assume that $\rho_0 \neq 0$. The behavior of $\mnDlogK(\rho)$ is then such that for $\rho$ small enough it is negative and $\mnDlogK(\rho \to 0) \to -\infty$, as depicted in Fig.~\ref{fig_inv_HO}. We therefore obtain three requirements for the function $F(\mnDlogK)$. First, $F(\mnDlogK \to -\infty) \to 0$ in order that the first term be regular. Second, $F(\mnDlogK) \leq 0$ for $\mnDlogK$ small enough in order for the first term to be non-negative. Third, $F^\prime(\mnDlogK) \geq 0$ in order for the second term to be non-negative. However, this is a contradiction, as these three requirements cannot be met simultaneously by a non-trivial $F(\mnDlogK)$. Accordingly, we conclude that only $\rho_0 = 0$ is viable.

%It therefore follows that only the case $\rho_0 = 0$ is viable. 
We therefore have
\begin{eqnarray}
\label{H0S1}
\mnDlogK &=& \coth\rho - \frac{1}{\rho}
\end{eqnarray}
and the general solution for the form factors read
%\begin{subequations}
%\begin{eqnarray}
%\label{Kthp}
%K &=& \frac{\rho}{\sinh\rho}
%\,,
%\\
%H &=&
%F_2^{-1}(\mnDlogK)
%\,.
%\end{eqnarray}
%\end{subequations}
\begin{equation}
\label{Kthp}
K \ = \ \frac{\rho}{\sinh\rho} \,,
\hspace{10mm}
H \ = \ F_2^{-1}(\mnDlogK) \,.
\end{equation}
Notice that for $F_2$ being the identity function this is nothing but the celebrated 't~Hooft--Polyakov solution in the BPS limit. We also remark that since $K = 1 - \frac{1}{6} \rho^2 + \mathcal{O}(\rho^4)$, the gauge fields are regular at the origin:
\begin{eqnarray}
A^i_a &=& - \frac{1}{6} \varepsilon_{abi} x_b v^2 g^2 \Big[1 + \mathcal{O}(\rho^2)\Big] \,,
\end{eqnarray}
as is customary to require.

For the energy density \eqref{EdensityInv} to be non-negative, the function $F$ must satisfy
\begin{equation}
F(\mnDlogK) \ \geq \ 0 \,,
\hspace{10mm}
F^\prime(\mnDlogK) \ \geq \ 0 \,,
\end{equation}
for all $\mnDlogK \in [0,1]$, while the requirement of regularity of $\mathcal{E}$ at $\rho = 0$ implies that $F(\mnDlogK \to 0)$ must go to zero sufficiently fast. For instance, if it behaves near the origin like a power function
\begin{eqnarray}
F(\mnDlogK) &\sim& \mnDlogK^N \,,
\end{eqnarray}
the exponent $N$ must satisfy
\begin{eqnarray}
\label{invNgeq1}
N &\geq& 1 \,.
\end{eqnarray}

We also see that, except in the special case $N = 1$ when $\mathcal{E}(0)/v^4g^2 = F^\prime(0)/3$, the energy density is always vanishing at $\rho=0$. In other words, most of the energy of the monopole is concentrated not in its centre, but rather in a spherical shell around it. This is very similar to the \emph{hollow monopoles} discussed in \cite{Bazeia:2018fhg}. However, while the authors of \cite{Bazeia:2018fhg} found and studied the hollow monopoles only numerically, we are going to present analytic solutions for them.

\section{Examples}
\label{examples}

By choosing $G(\mnDlogK) = \mnDlogK$ we have fulfilled the first point in our recipe of Sec.~\ref{cookbook}, so let us move to the next two points and provide some particular examples of functions $F(\mnDlogK)$ (to be able to draw energy density) and $F_2(H)$ (to be able to write down Lagrangian density.)
%Having introduced in the previous section the only known (at least to us) suitable function $G(\mnDlogK) = \mnDlogK$, let us now move on to the next two points from the recipe of Sec.~\ref{cookbook} and provide some particular examples of functions $F(\mnDlogK)$ (to be able to draw energy density) and $F_2(H)$ (to be able to write down Lagrangian density.)

\subsection{Power function}
\label{examplepower}

\begin{figure}[t]
\begin{center}
\includegraphics[width=0.5\textwidth]{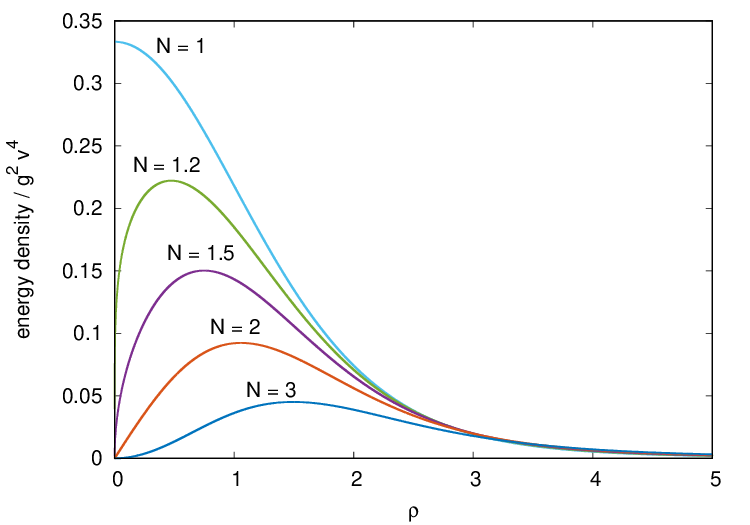}
\caption{Energy density for a single monopole solution of the power-function theory \eqref{lagrangianPowerInv} for various $N = n/m$.}
\label{plot_inv_pow_E}
\end{center}
\end{figure}

As the simplest example let us take
\begin{eqnarray}
F &=& \mnDlogK^N \,.
\end{eqnarray}
In order that the energy density (with $G(\mnDlogK) = \mnDlogK$ and, correspondingly, $K$ and $\mnDlogK$ given by \eqref{H0S1} and \eqref{Kthp})
%Taking further $G(\mnDlogK) = \mnDlogK$, we find that $N \geq 1$ in order that the corresponding energy density
\begin{subequations}
\begin{eqnarray}
\frac{\mathcal{E}}{v^4g^2}
&=&
\mnDlogK^{N} \bigg[2 \mnDlogK \frac{K^2}{\rho^2}
+ N \frac{1}{\mnDlogK} \frac{(1-K^2)^2}{\rho^4}\bigg]
\\
&\xrightarrow[\rho \to 0]{}& 
\rho^{N-1}
%\frac{N+2}{3^{N+1}}
\bigg[
\frac{N+2}{3^{N+1}}
%- \frac{(N+3)(N+4)}{5 \times 3^{N+2}} \rho^2
+ \mathcal{O}(\rho^2)
\bigg]
%\,.
%\nonumber \\ &&
\end{eqnarray}
\end{subequations}
be regular in the origin, we must demand $N\geq 1$ in accordance with \eqref{invNgeq1}. In Fig.~\ref{plot_inv_pow_E} we plot $\mathcal{E}$ for various values of $N$. As advertised, for $N > 1$ most of the energy is concentrated away from the centre and the monopoles are hollow.

In order to write down the corresponding Lagrangian, we choose, for simplicity, the function $F_2(H)$ to be a power function, too. Thus, we arrive at
\begin{eqnarray}
\label{lagrangianPowerInv}
\eL &=&
\frac{v^2}{2} H^{n+m} \bigg[
\frac{(D^\mu \threevector{\phi})^2}{\threevector{\phi}^2}
+ (nm-1) \frac{(\threevector{\phi} \innerdot D^\mu \threevector{\phi})^2}{\threevector{\phi}^4}
\bigg]
\nonumber \\ && {}
- \frac{1}{4g^2} H^{n-m} \bigg[
(\threevector{F}^{\mu\nu})^2
+ \bigg(\frac{n}{m}-1\bigg) \frac{(\threevector{\phi} \innerdot \threevector{F}^{\mu\nu} )^2}{\threevector{\phi}^2}
\bigg]
\,,
\hspace{8mm}
\end{eqnarray}
which in the language of $F_i$'s corresponds to
\begin{subequations}
\label{modelthooftpolyakov}
\begin{eqnarray}
F_1 = F_3 &=& H^n \,, \\
F_2 = F_4 &=& H^m \,,
\end{eqnarray}
\end{subequations}
provided that
% $F = \mnDlogK^{n/m}$.
\begin{eqnarray}
N &=& \frac{n}{m} \,.
\end{eqnarray}
Since $N \geq 1$, the parameters $n$, $m$ must therefore satisfy
\begin{eqnarray}
\label{nmgeq1}
\frac{n}{m} &\geq& 1 \,.
\end{eqnarray}
The solution follows immediately as
%\begin{subequations}
%\begin{eqnarray}
%K &=& \frac{\rho}{\sinh \rho} \,, \\
%\label{HpowerInv}
%H &=& \bigg(\coth\rho-\frac{1}{\rho}\bigg)^{\frac{1}{m}} \,.
%\end{eqnarray}
%\end{subequations}
\begin{equation}
\label{HpowerInv}
K \ = \ \frac{\rho}{\sinh \rho} \,,
\hspace{10mm}
H \ = \ \bigg(\coth\rho-\frac{1}{\rho}\bigg)^{\frac{1}{m}} \,.
\end{equation}

Interestingly, while the Lagrangian \eqref{lagrangianPowerInv} is form-invariant under general transformations $H \to \tilde{H}$ of Eq.~\eqref{transfomationPhiToTildePhi}, there is also a special subclass of these transformations that affects only the parameters $n$ and $m$.
%under which the Lagrangian is manifestly \emph{invariant}% form-invariance is in this case manifest.
Namely, upon transforming
\begin{eqnarray}
\label{transftHPH}
H &\to& H^\sigma \,,
\end{eqnarray}
where $\sigma \neq 0$ is arbitrary, we obtain a model of the same form as \eqref{lagrangianPowerInv}, up to the rescaling of parameters
\begin{equation}
\label{transftHPn12}
n \to \sigma n \,,
\hspace{10mm}
m \to \sigma m \,.
\end{equation}
Note that the physical (measurable) parameter $N = n/m$ is invariant under \eqref{transftHPn12}.
%Is is important to notice that the two models, with parameters $n$, $m$ and with $\sigma n$, $\sigma m$, are of course physically equivalent. This is exemplified by the fact that the physical (measurable) parameter $N = n/m$ is invariant under \eqref{transftHPn12}.

Obviously, for $n = m = 1$ this is just the well known 't~Hooft--Polyakov monopole in the renormalizable Georgi--Glashow model (in the BPS limit) \cite{tHooft:1974kcl,Polyakov:1974ek}. However, any model with $n/m = 1$ is equivalent to the Georgi--Glashow model, even though the Lagrangian and the solution $H$ are different.

%The form factor $H$, Eq.~\eqref{HpowerInv}, behaves in the origin like $H = (\rho/3 + \mathcal{O}(\rho^3))^{1/m}$ and, correspondingly, $\threevector{\phi} = v H \threevector{x}/r = v^2 g \threevector{x} \rho^{1/m-1} (1/3+\mathcal{O}(\rho^2))^{1/m}$. If $1/m \geq 1$, the field $\threevector{\phi}$ is regular, as is customary to require. If, on the other hand, $1/m < 1$, the field $\threevector{\phi}$ is singular at the origin. However, this singularity is completely unphysical and harmless, as it can be transformed away using \eqref{transftHPH} with, e.g., $\sigma = 1/m$.

%The form factor $H$, Eq.~\eqref{HpowerInv}, behaves in the origin like $H = (\rho/3 + \mathcal{O}(\rho^3))^{1/m}$ and, correspondingly, $\threevector{\phi} = v H \threevector{x}/r = v^2 g \threevector{x} \rho^{1/m-1} (1/3+\mathcal{O}(\rho^2))^{1/m}$. If $1/m \geq 1$, the field $\threevector{\phi}$ is regular, as is customary to require. If, on the other hand, $1/m < 1$, the field $\threevector{\phi}$ is singular at the origin. However, this singularity is completely unphysical and harmless, as it can be transformed away using \eqref{transftHPH}.
%% with, e.g., $\sigma = 1/m$.
%
%An especially convenient choice is in this regard to take $\sigma = 1/m$ in \eqref{transftHPH} which, indeed, renders $\threevector{\phi}$ to be regular in the origin (and, consequently, continuous and vanishing, as dictated by topological considerations). Moreover, it also makes the Lagrangian contain no negative powers of $H$ (due to $N = n/m \geq 1$) and thus well defined for all finite $H$.

There are now two issues. First, the form factor $H$, Eq.~\eqref{HpowerInv}, behaves in the origin like $H = (\rho/3 + \mathcal{O}(\rho^3))^{1/m}$ and, correspondingly, $\threevector{\phi} = v H \threevector{x}/r = v^2 g \threevector{x} \rho^{1/m-1} (1/3+\mathcal{O}(\rho^2))^{1/m}$. Thus, if $1/m < 1$, the field $\threevector{\phi}$ is singular in the origin. Second, the Lagrangian \eqref{lagrangianPowerInv} contains three different powers of $H$: $H^{n+m}$, $H^{n+m-2}$, $H^{n-m}$. Obviously, for some $n$, $m$ (even those satisfying \eqref{nmgeq1}) these powers can be negative and, consequently, singular for vanishing $H$. However, both issues can be cured by invoking the transformation \eqref{transfomationPhiToTildePhi}, namely its incarnation \eqref{transftHPn12} with
\begin{eqnarray}
\sigma &=& \frac{1}{m} \,.
\end{eqnarray}
%$\sigma = 1/m$.
This transformation leads to manifestly regular $\threevector{\phi} = v^2 g \threevector{x} (1/3+\mathcal{O}(\rho^2))$, while the potentially negative powers of $H$ in \eqref{lagrangianPowerInv} transform to $H^{\frac{n}{m}+1}$, $H^{\frac{n}{m}-1}$, $H^{\frac{n}{m}-1}$, that are already safely non-negative due to \eqref{nmgeq1}.

%+$n \to \frac{n}{m}$ and $m \to 1$

\subsection{Power-exponential function}

As a slightly more complicated example, let us consider %the previous power-function model modulated by an exponential function:
\begin{eqnarray}
F &=& \mnDlogK^N \exp\Big[a(\mnDlogK^M-1)^k\Big]
\,.
\end{eqnarray}
Here we assume $a \neq 0$ and $M \neq 0$, as otherwise we would obtain the previous case.

We take again $G = \mnDlogK$. From the requirements that $F(\mnDlogK = 1) = 1$ and $F^\prime = F [N + a k M \mnDlogK^M (\mnDlogK^M-1)^{k-1}] / \mnDlogK \geq 0$ we find that $N \geq 0$ and $aM > 0$, while $k$ must be a positive, odd integer. Now there are two options with rather different properties:
\begin{itemize}
\item $a > 0$ and $M > 0$: In this case the energy density behaves for small $\rho$ like a power function $\mathcal{E} \sim \rho^{N-1}$, so that it must be $N \geq 1$.

\item $a < 0$ and $M < 0$: In this case we have $\mathcal{E} \sim \exp(-1/\rho^{k|M|})$ (times a power function) as $\rho \to 0$ and $N \geq 0$.
\end{itemize}

\begin{figure}
\centering
\begin{minipage}[b]{.5\textwidth}
\centering
\includegraphics[width=1.0\textwidth]{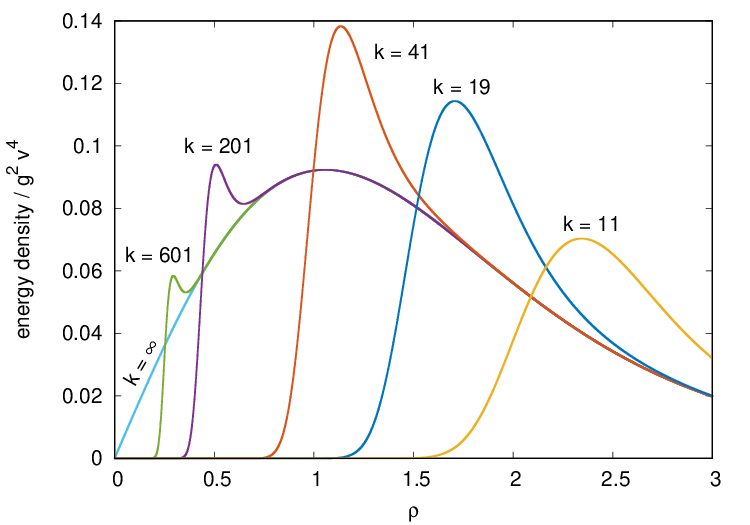}
\end{minipage}
%\begin{minipage}[b]{.5\textwidth}
%\centering
%\includegraphics[width=1.0\textwidth]{plot_inv_powexp_2min_2_E.eps}
%\end{minipage}
\caption{Energy density for a single monopole solution of the power-exponential-function theory \eqref{eq:lagg2} with $N = M = 2$,
% and either 
$a = 100$ and various $k$.
% (top panel) or $k = 501$ and various $a \geq 0$ (bottom panel).
}
\label{plot_inv_powexp_2minima}
\end{figure}

\begin{figure}[t]
\begin{center}
\includegraphics[width=0.5\textwidth]{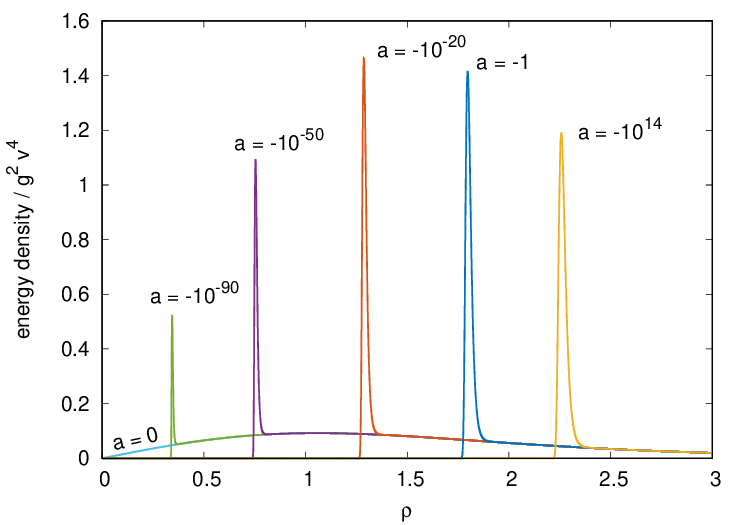}
\caption{Energy density for a single monopole solution of the power-exponential-function theory \eqref{eq:lagg2} with $N = 2$, $M = -1$, $k = 101$ and various $a \leq 0$.}
\label{plot_inv_powexp_expminimum}
\end{center}
\end{figure}

As an example of the Lagrangian that corresponds to this $F$, we can take
\begin{eqnarray}
\label{eq:lagg2}
\eL &=& 
\e^{a(H^{n\ell}-1)^k}H^{n-m} \nonumber \\
& & \times \Bigg\{
\frac{v^2}{2} H^{2m} \bigg[
\frac{(D^\mu \threevector{\phi})^2}{\threevector{\phi}^2}
+ \big(mn s -1\big) \frac{(\threevector{\phi} \innerdot D^\mu \threevector{\phi})^2}{\threevector{\phi}^4}
\bigg]
\nonumber \\ && {} \hspace{5mm}
- \frac{1}{4g^2} \bigg[
(\threevector{F}^{\mu\nu})^2
+ \bigg(\frac{n}{m}s-1\bigg) \frac{(\threevector{\phi} \innerdot \threevector{F}^{\mu\nu})^2}{\threevector{\phi}^2}
\bigg]
\Bigg\}
\,,
\hspace{8mm}
\end{eqnarray}
where we denoted for brevity
\begin{eqnarray}
s &\equiv& 1 + a\ell k\big(H^{n\ell}-1\big)^{k-1} H^{n\ell} \,.
\end{eqnarray}
In the language of $F_i$'s we would specify the theory by
\begin{subequations}
\begin{eqnarray}
F_1 = F_3 &=& H^n \exp\Big[a(H^{n\ell}-1)^k\Big] \,,
\\
F_2 = F_4 &=& H^{m} \,,
\end{eqnarray}
\end{subequations}
which is related to the $F$ by the identification
\begin{equation}
N \ = \ \frac{n}{m} \,,
\hspace{10mm}
M \ = \ \ell\frac{n}{m} \,.
\end{equation}
The form factors $K$ and $H$ are given by the very same expressions \eqref{HpowerInv} as in the previous power-function model.

%The form factors (again) read
%%\begin{subequations}
%%\begin{eqnarray}
%%K &=& \frac{\rho}{\sinh \rho} \,, \\
%%H &=& \bigg( \coth\rho - \frac{1}{\rho} \bigg)^{\frac{1}{m}} \,.
%%\end{eqnarray}
%%\end{subequations}
%\begin{equation}
%K \ = \ \frac{\rho}{\sinh \rho} \,,
%\hspace{10mm}
%H \ = \ \bigg(\coth\rho-\frac{1}{\rho}\bigg)^{\frac{1}{m}} \,.
%\end{equation}

The reason to consider this complicated and rather artificial model is that it showcases more interesting energy density profiles with a richer structure (admittedly, for \qm{unnaturally} large values of parameters). While we again obtain the hollow monopoles, this time they have new features. As depicted in Fig.~\ref{plot_inv_powexp_2minima}, for some range of parameters there can be not only one minimum of the energy density, but two. In Fig.~\ref{plot_inv_powexp_expminimum}, we see that the peak of the energy density can be much sharper. Moreover, since in the latter case the energy density falls off exponentially for small $\rho$ (due to $a < 0$), there is a well-defined finite region in the centre of the monopole with virtually vanishing energy density -- accordingly we can dub these solutions the \emph{truly hollow} monopoles.

\section{Summary and outlook}
\label{conclusions}

In this paper we have investigated spherically symmetric solutions of a family of general $SU(2)$ gauge theories \eqref{lagrangian} with adjoint scalar in the BPS limit. 

First, we discussed the redundancy and physical r\^{o}le of the four functions $f_i^2$ that define our model. %The with non-canonical kinetic terms controlled by four form-functions $f_i^2(H)$.%$ in the BPS limit. 
There is a form-invariance of our theory \eqref{lagrangian} that exploits a general field redefinition \eqref{transfomationPhiToTildePhi} of the gauge-invariant part of the adjoint scalar $\threevector{\phi}$. This redefinition leaves the structure of the model intact and allows to fix one of the four functions (e.g., to have the kinetic term of $\threevector{\phi}$ canonically normalized). By itself, however, the transformation \eqref{transfomationPhiToTildePhi} is not enough to eliminate non-canonical structure of the (dressed) gauge fields. To achieve that, we have introduced a form non-invariant transformation \eqref{transAtotildeAspec} that defines canonically normalized gauge fields and through which the starting Lagrangian \eqref{lagrangian} is recast into \eqref{tildeLafteronlyA}. In this way, the presence of field-dependent $SU(2)$ magnetic susceptibility (functions $f_2^2$ and $f_4^2$) in Eq.~\eqref{lagrangian} are found in Eq.~\eqref{tildeLafteronlyA} to control the strength of dipole-moment interactions. This clearly illustrates the physical r\^{o}le of $f_2^2$ and $f_4^2$ and opens up an avenue for further investigations of their impact on physics that we plan to do in the future. (In App.~\ref{AppendixFieldRedef}, we discuss a generalization of both transformations and showcase its group-like properties and the impact on the hedgehog Ansatz).

Let us also stress that the transformations \eqref{transfomationPhiToTildePhi} and \eqref{transAtotildeAspec} will generally change the path-integral measure.
Therefore, while the transformed models \eqref{lagrangianTransf} and \eqref{tildeLafteronlyA} are classically equivalent to the original model \eqref{lagrangian}, at the quantum level their relations may be more complicated.
%so that the relations between dressed Eq.~\eqref{lagrangian} and canonical models Eq.~\eqref{tildeLafteronlyA}, which are equivalent at the classical level, may be more complicated at the quantum level.
Although far outside of the scope of this paper, it would be interesting to see if there is a subset of the general transformations exposed in App.~\ref{AppendixFieldRedef} that preserve the path-integral measure,
so that the transformed mo\-dels could be equated even at the quantum level.
%so that we can equate canonical and dressed models even at the quantum level.

Second, we have formulated a BPS version of the theory with the help of reparametrizing the Lagrangian \eqref{lagrangian} in terms of new, more \qm{BPS-friendly} functions $F_i$ (rather than $f_i^2$). This has several advantages: (i) the key BPS condition is stated very simply as $F_3 = F_1$, (ii) the form-invariance \eqref{transfomationPhiToTildePhi} becomes a simple function composition, and (iii) the solvable cases that are studied in this paper -- all of which can be regarded as certain generalizations of the 't~Hooft--Polyakov monopole -- belong to a special subset that is defined simply as $F_4 = F_2$.

Third, in the BPS limit there are only two physically relevant form-functions that we have labelled $F$ and $G$ (see Eq.~\eqref{eq:fandg}). We have then constructed an especially convenient form of the Lagrangian, \eqref{lagrangianFG}, which is written solely in terms of form-functions $F$, $G$ and $F_2$, with the freedom of performing \eqref{transfomationPhiToTildePhi} being entirely contained in $F_2$. Indeed, this function does not appear in a general formula \eqref{EdensityInv} for energy density of spherically symmetric solutions, which is thus manifestly form-invariant.

Having clarified the r\^{o}le of various functions, the rest of the paper presents in Sec.~\ref{examples} some concrete examples of analytic BPS solutions that illustrate an application of a general \qm{recipe} described in Sec.~\ref{cookbook}. These examples are not meant to be exhaustive, rather they expose a key point of this study: How the shape of a monopole (i.e., distribution of its energy) depends on the form-function $F$. (The other form-function, $G$, is being fixed for simplicity, although we suspect its r\^{o}le to be qualitatively the same.)
%In the examples presented -- and with the general recipe of Sec.~\ref{cookbook} in mind -- we can conclude that `physical' form-functions $F$ and $G$ controls to a large extend the shape of monopole -- i.e. the profile of energy density as can be seen from Eq.~\eqref{EdensityInv}. 
While the general dependence might be glimpsed from the formula \eqref{EdensityInv}, in particular, we have seen that the position and number of extrema can be controlled by the choice of $F$ (see Fig.~\ref{plot_inv_powexp_2minima}).

A universal feature seems to be the tendency of energy to concentrate in a shell rather than at the monopole's centre. In other words, a generic monopole of our family of models tends to be \emph{hollow}. The only exception in the examples presented here are the cases $n = m$ of the simplest power-function model \eqref{lagrangianPowerInv} that are actually nothing but the 't~Hooft--Polyakov monopole, up to a field redefinition \eqref{transfomationPhiToTildePhi}. These cases also stand out by the presence of the canonically normalized gauge kinetic term in the Lagrangian density.

On the other hand, in all our examples of hollow monopoles the gauge kinetic terms are always modified by some power of the scalar singlet $H$, manifesting a non-trivial $SU(2)$ magnetic dipole-moment of the adjoint triplet. Intuitively, the presence of hollow monopoles therefore seems to be connected with vanishing (or \qm{freezing}) of the gauge kinetic term at the monopole's centre, namely that the dipole-moment interactions that \emph{screen} the bare monopole charge become effectively infinite. At this point, we present it as an observation, however, we are planing to examine this issue more thoroughly by investigating the properties of homogeneous \qm{phases} of the theory \eqref{lagrangian} in a future work.

Lastly, in this paper, we have only expounded the case of \emph{invertible} $F_2$ that allowed us to condense the system of two first-order BPS equations \eqref{logKH} into a single second-order equation \eqref{eqKinS}. However, there is an entire branch of analytic solutions that correspond to a \emph{non-invertible} choices of $F_2$. For instance, taking $F_2 = 1$ yields a particularly simple form of $K$:
%(while $H$, which depends on other form-functions, being typically much more complicated):
\begin{eqnarray}
K &=& \xi \e^{-\rho} \,,
\end{eqnarray} 
where $\xi$ is a constant of integration. (The other form factor, $H$, depends on $F_4$ and is typically much more complicated.) First of all, near origin $K \sim \xi (1-\rho + \mathcal{O}(\rho^2))$ and, correspondingly, $A_a^i \sim \varepsilon_{abi}x_b (\xi - 1 + \mathcal{O}(r))/r^2$ as~$r\to 0$. Notice that this singularity only becomes milder if $\xi = 1$, but does not disappear completely. As discussed at the beginning of this paper, by itself a singular behavior might not be problematic, as it can be transformed away by a general field redefinition of the type discussed in Appendix~\ref{AppendixFieldRedef}. Interestingly, however, the physical requirements (like regularity of the energy density) do not constrain $\xi$ to have a single value, but rather $\xi \in [-1,1]$. In other words, $\xi$ is a completely new moduli of the BPS solution and it is physical in the sense that it controls the shape of the energy density. In particular, it is related to the width of the hollow cavity.

We plan to investigate this class of solutions in a separate paper. There we hope to elucidate the physical origin of this moduli and whether the corresponding solutions are physically viable. In particular, we have to analyze the stability of these solutions, as the presence of $\xi$ might lead to dynamical instabilities. Further, we need to study them in non-BPS cases and identify what are the exact conditions that lead to their presence.

\begin{acknowledgements}
The authors would like to express the gratitude for the institutional support of the Institute of Experimental and Applied Physics, Czech Technical University in Prague (P.~B.~and F.~B.), and of the Research Centre for Theoretical Physics and Astrophysics, Institute of Physics, Silesian University in Opava (F.~B.).
%F. B. would like to express gratitude for the institutional support of the Research Centre for Theoretical Physics and Astrophysics, Institute of Physics, Silesian University in Opava and to the Institute of Experimental and Applied Physics, Czech Technical University in Prague.
P.~B.~is indebted to A\v{s}tar \v{S}eran and FSM for invaluable discussions.
\end{acknowledgements}

\appendix

\section{General canonical transformation for $SU(2)$ adjoint fields}
\label{AppendixFieldRedef}

%\subsection{The transformation}

Let us consider the transformation $\{\threevector{A}_\mu, \threevector{\phi}\} \to \{\tilde{\threevector{A}}_\mu, \tilde{\threevector{\phi}}\}$ generated by the functions $\HtoH$, $h$, $k$, $\ell$ as
\begin{subequations}
\label{transformationgeneral}
\begin{eqnarray}
\label{transformationgeneralphi}
\threevector{\phi} &=& 
\frac{\HtoH(\tilde{H})}{\tilde{H}} \tilde{\threevector{\phi}}
\,,
%\end{eqnarray}
%\begin{eqnarray}
\\
\threevector{A}_\mu &=&
\tilde{\threevector{A}}_\mu
+ h(\tilde{H})
\smash{
\overbrace{
\bigg[ \frac{\tilde{D}_\mu \tilde{\threevector{\phi}}}{v\tilde{H}}
- \frac{\tilde{\threevector{\phi}} \innerdot \tilde{D}_\mu \tilde{\threevector{\phi}}}{(v\tilde{H})^3} \tilde{\threevector{\phi}} \bigg]
}^{
\mathclap{
\displaystyle
\tilde{D}_\mu \threevector{n}
}}
}
+ k(\tilde{H}) \frac{\tilde{\threevector{\phi}} \innerdot \tilde{D}_\mu \tilde{\threevector{\phi}}}{(v\tilde{H})^3} \tilde{\threevector{\phi}}
\nonumber \\ && {}
+ \big[\ell(\tilde{H})-1\big] \frac{\tilde{\threevector{\phi}} \times \tilde{D}_\mu \tilde{\threevector{\phi}}}{(v\tilde{H})^2}
\,,
\end{eqnarray}
\end{subequations}
where $\tilde{\threevector{\phi}} = v \tilde{H} \threevector{n}$ and $\tilde{D}_\mu \tilde{\threevector{\phi}} \equiv \partial_\mu \tilde{\threevector{\phi}} + \tilde{\threevector{A}}_\mu \times \tilde{\threevector{\phi}}$. Equation \eqref{transformationgeneralphi} is equivalent to $H = \HtoH(\tilde H)$.

The transformations of the covariant derivative and field-strength tensor follow as
\begin{eqnarray}
\frac{D_\mu \threevector{\phi}}{vH} &=& \ell \frac{\tilde{D}_\mu \tilde{\threevector{\phi}}}{v\tilde{H}}
- h \frac{\tilde{\threevector{\phi}} \times \tilde{D}_\mu \tilde{\threevector{\phi}}}{(v\tilde{H})^2}
- \bigg(\ell - \tilde{H}\frac{\HtoH^\prime}{\HtoH} \bigg) \frac{\tilde{\threevector{\phi}} \innerdot \tilde{D}_\mu \tilde{\threevector{\phi}}}{(v\tilde{H})^3} \tilde{\threevector{\phi}}
\,,
\nonumber \\ && {}
\\
\threevector{F}_{\mu\nu} &=&
\ell \, \tilde{\threevector{F}}_{\mu\nu}
- h \frac{\tilde{\threevector{\phi}} \times \tilde{\threevector{F}}_{\mu\nu}}{v\tilde{H}}
- \big(\ell-1\big) \frac{\tilde{\threevector{\phi}} \cdot \tilde{\threevector{F}}_{\mu\nu}}{(v\tilde{H})^2} \tilde{\threevector{\phi}}
+ \threevector{d}_{\mu\nu}
\,,
\nonumber \\ &&
\end{eqnarray}
where we defined
\begin{widetext}
\begin{eqnarray}
\threevector{d}_{\mu\nu} &\equiv&
\phantom{+\,\,}
\big(\tilde{H}h^{\prime} - k \ell\big) \frac{(\tilde{\threevector{\phi}} \innerdot \tilde{D}_\mu\tilde{\threevector{\phi}})(\tilde{D}_\nu\tilde{\threevector{\phi}}) - (\mu\leftrightarrow\nu)}{(v\tilde{H})^3}
%\nonumber \\ && {}
+ \big(\tilde{H}\ell^{\prime}+kh\big) \frac{(\tilde{\threevector{\phi}} \innerdot \tilde{D}_\mu\tilde{\threevector{\phi}})(\tilde{\threevector{\phi}} \times \tilde{D}_\nu\tilde{\threevector{\phi}}) - (\mu\leftrightarrow\nu)}{(v\tilde{H})^4}
\\ \nonumber && {}
- \Big[(\ell^2+h^2-1)-(\ell-1)^2\Big] \bigg[
%\underbrace{
%\big(2(\ell-1)+h^2\big)
%}_{\mathclap{\displaystyle
%(\ell^2+h^2-1)-(\ell-1)^2
%}}
\frac{(\tilde{\threevector{\phi}} \innerdot \tilde{D}_\mu\tilde{\threevector{\phi}})(\tilde{\threevector{\phi}} \times \tilde{D}_\nu\tilde{\threevector{\phi}}) - (\mu\leftrightarrow\nu)}{(v\tilde{H})^4}
%\nonumber \\ && {}
-
%\overbrace{
%\big(2(\ell-1)+h^2\big)
%}^{}
\frac{\tilde{D}_\mu\tilde{\threevector{\phi}} \times \tilde{D}_\nu\tilde{\threevector{\phi}}}{(v\tilde{H})^2}
\bigg]
%\nonumber \\ && {}
+ (\ell-1)^2 \frac{\tilde{\threevector{\phi}} \innerdot (\tilde{D}_\mu\tilde{\threevector{\phi}} \times \tilde{D}_\nu\tilde{\threevector{\phi}})}{(v\tilde{H})^4} \tilde{\threevector{\phi}}
\,,
%\nonumber \\ && {}
\end{eqnarray}
%\end{widetext}
with the primes being differentiations with respect to $\tilde{H}$. Under the transformation \eqref{transformationgeneral} the original Lagrangian $\eL$, Eq.~\eqref{lagrangian}, transforms to another Lagrangian $\tilde\eL$ of the form
%\begin{widetext}
\begin{eqnarray}
\tilde\eL &=&  
\frac{v^2}{2} \bigg[
\tilde f_1^2 \bigg( \frac{(\tilde D^\mu \tilde{\threevector{\phi}})^2}{\tilde{\threevector{\phi}}^2} - \frac{(\tilde{ \threevector{\phi}} \innerdot \tilde D^\mu \tilde{\threevector{\phi}})^2}{\tilde{ \threevector{\phi}}^4}\bigg)
+ \tilde f_3^2 \frac{(\tilde{\threevector{\phi}} \innerdot \tilde D^\mu \tilde{\threevector{\phi}})^2}{\tilde{ \threevector{\phi}}^4}
\bigg]
%\nonumber \\ && {}
- \frac{1}{4g^2} \bigg[
\tilde f_2^2 \bigg((\threevectorRedef{F}^{\mu\nu})^2 - \frac{(\tilde{ \threevector{\phi}} \innerdot \threevectorRedef{F}^{\mu\nu})^2}{\tilde{ \threevector{\phi}}^2}\bigg)
+ \tilde f_4^2 \frac{(\tilde{\threevector{\phi}} \innerdot \threevectorRedef{F}^{\mu\nu})^2}{\tilde{\threevector{\phi}}^2}
\bigg]
%\nonumber \\ && {}
%- \tilde V(\tilde{\threevector{\phi}}^2)
\nonumber \\ && {}
- \frac{1}{2g^2} \threevector{d}^{\mu\nu} \innerdot \Bigg\{
f_2^2 \ell \bigg[ 
\tilde{\threevector{F}}_{\mu\nu}
- \frac{(\tilde{\threevector{\phi}} \innerdot \tilde{\threevector{F}}_{\mu\nu})}{\tilde{\threevector{\phi}}^2} \tilde{\threevector{\phi}}
\bigg]
%\nonumber \\ && {} \hspace{14mm}
+ f_2^2 h \frac{\tilde{\threevector{F}}_{\mu\nu} \times \tilde{\threevector{\phi}}}{v\tilde{H}}
%\nonumber \\ && {}
+ f_4^2 \frac{(\tilde{\threevector{\phi}} \innerdot \tilde{\threevector{F}}_{\mu\nu})}{\tilde{\threevector{\phi}}^2} \tilde{\threevector{\phi}}
\Bigg\}
\nonumber \\ && {}
- \frac{1}{4g^2} \bigg[
f_2^2 \bigg((\threevector{d}_{\mu\nu})^2 - \frac{(\tilde{\threevector{\phi}} \innerdot \threevector{d}_{\mu\nu})^2}{\tilde{\threevector{\phi}}^2}\bigg)
%\nonumber \\ && {}
+ f_4^2 \frac{(\tilde{\threevector{\phi}} \innerdot \threevector{d}_{\mu\nu})^2}{\tilde{\threevector{\phi}}^2}
\bigg]
%\nonumber \\ && {}
- \tilde V(\tilde{\threevector{\phi}}^2)
\,,
\end{eqnarray}
\end{widetext}
%\remark{Typo: correction on the first line: from $\threevector{F}$ to $\tilde{\threevector{F}}$}
where the new functions $\tilde f_i^2 = \tilde f_i^2(\tilde{H})$ are given as
\begin{subequations}
\begin{eqnarray}
\tilde f_i^2 &=& f_i^2 \big(\ell^2+h^2\big) \,,
\hspace{6mm}
(i=1,2)
\\
\tilde f_3^2 &=& f_3^2 \bigg(\tilde{H}\frac{\HtoH^\prime}{\HtoH}\bigg)^2 \,,
\\
\tilde f_4^2 &=& f_4^2 \,,
\end{eqnarray}
\end{subequations}
with $f_i^2 = f_i^2(\HtoH(\tilde{H}))$, and $\tilde V(\tilde{\threevector{\phi}}^2) = V(\threevector{\phi}^2)$.

We also obtain
\begin{eqnarray}
\threevector{\phi} \innerdot \threevector{A}_\mu &=&
\frac{\HtoH}{\tilde{H}} \bigg[
\tilde{\threevector{\phi}} \innerdot \tilde{\threevector{A}}_\mu
+ k \frac{\tilde{\threevector{\phi}} \innerdot \tilde{D}_\mu \tilde{\threevector{\phi}}}{v\tilde{H}}
\bigg]
\,,
\hspace{20mm}
\\
\threevector{\phi} \innerdot \threevector{F}_{\mu\nu} &=&
\frac{\HtoH}{\tilde{H}}\bigg[
\tilde{\threevector{\phi}} \innerdot \tilde{\threevector{F}}_{\mu\nu}
+ 
%\underbrace{
%\tilde{\threevector{\phi}} \innerdot \threevector{d}_{\mu\nu}
%}_{\mathclap{\displaystyle
%\hspace{10mm}
(\ell^2+h^2-1) \frac{\tilde{\threevector{\phi}} \innerdot (\tilde{D}_\mu\tilde{\threevector{\phi}} \times \tilde{D}_\nu\tilde{\threevector{\phi}})}{\tilde{\threevector{\phi}}^2}
%}}
\bigg]
\,.
\nonumber \\ &&
\end{eqnarray}

%\subsection{Group law}

By performing two consecutive transformations \eqref{transformationgeneral} we obtain again a transformation of the form \eqref{transformationgeneral}. For the scalar it is obvious: If we transform first $H_1 \to H_2$ and then $H_2 \to H_3$ with $H_1 = \HtoH_2(H_2)$ and $H_2 = \HtoH_3(H_3)$, respectively, then the combined transformation $H_1 \to H_3$ is done via $H_1 = \HtoH_2(\HtoH_3(H_3))$, i.e., by simple function composition. For the gauge fields the combined transformation is more complicated. If the first transformation is by $h_2(H_2)$, $k_2(H_2)$, $\ell_2(H_2)$ and the second one by $h_3(H_3)$, $k_3(H_3)$, $\ell_3(H_3)$, then the combination of the two transformation is done via
\begin{subequations}
\begin{eqnarray}
h(H_3) &=& \ell_2(H_2) \, h_3(H_3) + h_2(H_2) \, \ell_3(H_3) \,,
\\
\ell(H_3) &=& \ell_2(H_2) \, \ell_3(H_3) - h_2(H_2) \, h_3(H_3) \,,
\\
k(H_3) &=& k_3(H_3) + k_2(H_2) \, H_3 \frac{\HtoH_3^\prime(H_3)}{\HtoH_3(H_3)} \,.
\hspace{10mm}
\end{eqnarray}
\end{subequations}
%where $H_2 = \HtoH_3(H_3)$.
%\remark{Typo: arguments on the l.h.s. corrected from $h_3$ to $H_3$}

%\subsection{Impact on the hedgehog Ansatz}

Unless $h = k = 0$, the general transformation \eqref{transformationgeneral} does \emph{not} protect the spherically symmetric Ansatz \eqref{ansatzcomponents}, but rather leads to
\begin{subequations}
\begin{eqnarray}
\tilde\phi_a &=& v \tilde{H} \frac{x_a}{r} \,,
\\
\tilde{A}^i_a &=& -\frac{\varepsilon_{abi} x_b}{r^2} (1-\tilde{K}) + \frac{\delta_{ia}r^2-x_ix_a}{r^3} \tilde{L} + \frac{x_ix_a}{r^3} \tilde{M} \,,
\hspace{11mm}
\end{eqnarray}
\end{subequations}
where $\tilde{H}$ is obtained by inverting $\HtoH(\tilde{H}) = H$, while
\begin{equation}
\tilde{K} \ \equiv\ \frac{\ell}{\ell^2+h^2} K \,,
\hspace{3mm}
\tilde{L} \ \equiv\ \frac{-h}{\ell^2+h^2} K \,,
\hspace{3mm}
\tilde{M} \ \equiv\ - k r \frac{\tilde{H}^\prime}{\tilde{H}} \,.
\end{equation}
%Notice that $\tilde{K}^2 + \tilde{L}^2 = K^2$.
Nevertheless,
%Although the general transformation \eqref{transformationgeneral} does not protect the form of the Ansatz \eqref{ansatzcomponents},
it is still possible to require the transformation \eqref{transformationgeneral} to protect, at least, the transversality of the gauge and scalar fields: $\threevector{\phi} \cdot \threevector{A}^i = \tilde{\threevector{\phi}} \cdot \threevectorRedef{A}^i = 0$. In other words, one can demand $\tilde{M} = 0$, implying $k = 0$.

%\bibliographystyle{JHEP}
%\bibliography{references}

\providecommand{\href}[2]{#2}\begingroup\raggedright\endgroup

\end{document}